\documentclass[twocolumn,showpacs,preprintnumbers,prd,superscriptaddress,nofootinbib]{revtex4-1}
\usepackage{amsmath}
\usepackage{graphicx}
\usepackage[titletoc]{appendix}
\usepackage[hang,normal]{subfigure}
\usepackage{color}

\newcommand{\tabref}[1]{Tab.~\ref{#1}}
\newcommand{\figref}[1]{Fig.~\ref{#1}}

\newcommand{\eqeqref}[1]{Eq.~\eqref{#1}}
\newcommand{\eqsref}[1]{Eqs.~\eqref{#1}}
\newcommand{\secref}[1]{Section~\ref{#1}}

\newcommand{\Tr}[0]{\mathrm{Tr}}
\newcommand{\mb}[1]{\mathbf{#1}}

\begin{document}

\title{A theory of finite-temperature Bose-Einstein condensates in neutron stars}

\author{Christine Gruber}
\email{christine.gruber@uni-oldenburg.de}
\affiliation{Institut f\"ur Theoretische Physik, Freie Universit\"at Berlin, Arnimallee 14, 14195 Berlin, Germany,}
\affiliation{Institut f\"ur Physik, Universit\"at Oldenburg, Carl-von-Ossietzky-Str. 9-11, 26129 Oldenburg, Germany,}

\author{Axel Pelster}
\email{axel.pelster@physik.uni-kl.de}
\affiliation{Fachbereich Physik und Forschungszentrum OPTIMAS, Technische Universit\"at Kaiserslautern, 
Erwin-Schr\"odinger-Strasse 46, 67663 Kaiserslautern, Germany.}

\date{\today}

\begin{abstract}
We investigate the possible occurrence of a Bose-Einstein condensed phase of matter within neutron stars 
due to the formation of Cooper pairs among the superfluid neutrons. To this end we study the 
condensation of bosonic particles under the influence of both a short-range contact and a long-range gravitational 
interaction in the framework of a  Hartree-Fock theory. We consider a finite-temperature scenario, 
generalizing existing approaches, and derive macroscopic and astrophysically relevant quantities like a mass 
limit for neutron stars. 
\end{abstract}

\maketitle
 
\section{Bose-Einstein condensates in Astrophysics}
\label{sec:BECintro}
In this work we present a model for a quantum phenomenon with impact on macroscopically large scales by considering 
the possible occurrence of a Bose-Einstein condensate (BEC) in compact astrophysical objects. Laboratory 
experiments on cold gases have first confirmed \cite{1995Corn,1995Kett} the existence of a particular state 
of matter for bosonic particles when cooled down to ultracold temperatures in low-density environments. 
Originating from Bose's re-derivation of Planck's law of black body radiation \cite{1924Bose}, Einstein predicted 
this phenomenon employing a new statistics for the distributions of massive bosons in an ensemble, thereby 
describing a synchronization of the wave functions of all particles in the system \cite{1925Eins}. 
Velocity-distribution data from experiments show a macroscopic occupation of the ground state, thus 
demonstrating the existence of a quantum phenomenon with impacts on large scales. \\
Even though the effect is known from laboratory physics, it can be considered in completely 
different circumstances as well, as for example in compact objects in astrophysics. Generally a 
BEC is created when the temperature in a system falls below the critical temperature 
\begin{equation} \label{eq:Tcrit}
  T_{\mathrm{crit}} = \left[ \frac{n}{\zeta(3/2)} \right]^{2/3} \frac{2\pi \hbar^2}{mk_B} \,,
\end{equation}
corresponding to the point where the thermal de Broglie wavelength equals the average interparticle 
distance, so the wave functions of individual particles overlap and synchronize. Rather surprisingly, 
considering the typical temperatures and densities in astrophysical scenarios extracted from observations, 
condition~\eqref{eq:Tcrit} seems to be met in some cases of compact objects. A possible example for BECs in 
compact objects in astrophysics are boson stars - either as an abstract concept of a bosonic field in a 
spherically symmetric metric \cite{2012Klei}, or as the concrete case of a star consisting of bosonic particles. 
Helium white dwarfs have been considered as candidates before \cite{2011Benv,2000NagC}, even though due to the 
ongoing fusion processes inside the star the abundance of objects solely made up of helium is presumably small. 
Another problem is posed by the ionization of Helium at temperatures higher than 
about $10^5\, \mathrm{K}$, which makes the theory of a BEC of neutral bosons effectively inapplicable in that case. 
More realistically, white dwarfs can be described by an approach considering a background lattice of positive ions immersed 
in a sea of electrons. \\
Alternatively, the existence of BECs in neutron stars has been suggested \cite{2011Cha2}. Neutron stars have been considered 
firstly by Tolman \cite{1939Tolm} as well as Oppenheimer and Volkoff \cite{1939Oppe}. They investigated a fluid of self-gravitating 
neutrons, for which the equation of state is determined by Fermi statistics, in the context of general relativity embedded in a 
spherically symmetric metric, and searched for stable equilibrium 
configurations of the system. In the scenario assumed by Tolman, Oppenheimer and Volkoff (TOV), the gravitational collapse 
of a cloud of neutrons is counterbalanced by the degeneracy pressure of the neutrons as a consequence of the Pauli 
exclusion principle. The maximum stable mass of such a system, the TOV limit, was found to be about $0.7 \,M_{\odot}$ 
\cite{1939Tolm,1939Oppe}. In contradiction to this original prediction, observations \cite{2010Demo} have found neutron stars with 
masses up to a value of $2\,M_{\odot}$. Hence, there has been an abundance of proposals and models to explain the observed masses 
of neutron stars \cite{2007Haen}. The existence of all kinds of states or types of matter in the core of the objects was suggested, 
reaching from strange baryons over heavy mesons like kaons or pions to quark matter, while the crust of neutron stars is usually 
assumed to consist of neutrons and electrons \cite{2012Belv}. \\
BECs in neutron stars are feasible despite the fact that neutrons are fermions. A general consensus exists over the 
fact that neutrons in a neutron star should be in a superfluid phase \cite{2011Page}, i.e. the particles are bound 
in Cooper pairs and can be treated as composite bosons with an effective mass of $m=2m_n$, which can form a BEC. 
A microscopically exact way of treating such a system is provided by the theory of a BCS-BEC-crossover \cite{1993Sade,1997Enge}, 
i.e. a transition from the quantum state of superfluidity (BCS phase) to a Bose-Einstein condensate. 
The theory describes the pairing mechanism between neutrons, allowing for a coexistence of single neutrons and 
neutron pairs in a mixed state of Fermi and Bose fluids. The phenomenon 
has been observed in the laboratory on weakly bound molecules formed by two fermionic atoms \cite{2003Grei}, and has 
more recently also been applied to the case of nuclear or neutron matter. Calculations in Refs. \cite{2005Astr,2006Mats,2007Marg,2013Sala} 
show that nuclear forces between nucleons, in particular neutrons, lead to the formation of nucleon pairs, which can be treated 
as effective bosons in a BEC under appropriate conditions. The phenomenon of nucleon pairing was firstly proposed in 1935 
by a phenomenological formula by Weizs\"acker \cite{1935Weiz} in the context of atomic nuclei. Later on, the superfluidity 
of fermionic particles was formulated microscopically exactly in terms of a BCS-type theory, which was then applied to the 
case of nucleons inside an atomic core, and by now the treatment of superfluidity in nuclear matter is well established \cite{2010Brin}. 
Superfluidity in the context of neutron stars can be described in the same way as in atomic nuclei - physically, neutron stars 
are nothing but a gigantic atomic nucleus, consisting of neutrons and protons which are subject to the same pairing effect as 
in atomic nuclei. \\
In the present work, we use several assumptions and simplifications which differ from the picture of an atomic 
nucleus. Firstly, we assume the system to be purely made up by neutrons, and neglect the presence of other particles 
as protons and electrons. Further, we approach the system in a purely phenomenological way and treat the paired neutrons 
as effective bosons which form the BEC. There is no fermionic component in our system, i.e. we assume the pairing of the 
neutrons as strong enough to be able to consider them as perfectly bosonic. Typical densities in the center of neutron stars 
lie around $10^{14} \,\mathrm{g}/\mathrm{cm}^3$, whereas in the outer regimes densities decrease to about $10^{6} \,\mathrm{g}/\mathrm{cm}^3$. 
Assuming an effective boson mass of $m=2m_n$, according to~\eqeqref{eq:Tcrit} this corresponds to critical temperatures of 
$10^{10}\, \mathrm{K}$ to $10^{5}\, \mathrm{K}$, respectively. Thus it is potentially possible during the initial stages of the 
evolution of a neutron star to fulfill condition~\eqref{eq:Tcrit} and consider the presence of a neutron-pair BEC. \\
Given that the known scattering length of neutrons in nuclear matter is quite large, the interior of neutron stars
is actually better described by the unitary regime, i.e. the transition phase between the BCS and the BEC limits. It is clear 
that for a realistic description it is necessary to consider also the single neutrons in the star and set up the exact theory 
of the BCS-BEC crossover. The cases of a pure BCS phase and a pure BEC phase then have to result as limits of this general 
crossover theory. In the literature, neutron stars are usually described in one of the limiting states, i.e. the BCS fluid. 
In this work we will investigate the opposite limit of a BEC fluid as a first step towards the unifying crossover theory.  \\
~\\
Systems of self-gravitating bosonic (and fermionic) particles have already been considered some time ago in Ref. \cite{1969Ruff}. 
For the case of Newtonian gravity, the investigations have resulted in unstable configurations for bosons, which could only 
be stabilized by the inclusion of general relativistic effects. However, in contrast to our model the particles 
in Ref. \cite{1969Ruff} are assumed to be free, only subject to gravitational interactions. In our model, 
contact interaction, i.e. hard shell scattering between bosons, will be employed to stabilize the 
system against gravitational collapse. Thus, even for zero temperature with vanishing thermal pressure and in 
the case of Newtonian gravity, contact interaction provides the necessary pressure to counterbalance gravity. \\
A system of bosons in a Bose-Einstein condensed phase with contact and gravitational interactions, such as the system 
we are considering, for the case of 
zero temperatures has recently been treated in Ref. \cite{2011Cha2} and applied to the example of superfluid neutron stars. 
A generalization to a BEC at finite temperatures was recently worked out in Ref. \cite{2012Hark}, but then 
applied to the example of a dark matter BEC in a Friedmann-Robertson-Walker universe. The theory of Bose-Einstein 
condensation for the case of bosonic dark matter was also considered by other authors, see 
Refs. \cite{2011Cha1,2011Hark,2012Liet}. Due to the widely unknown nature and properties of dark matter, it is, however, 
a rather speculative field, and the effects of the presence of a Bose-Einstein condensate of dark matter particles in 
contrast to thermal phase dark matter are difficult to detect, most likely only by the gravitational lensing behaviour 
of dark matter halos. The environmental conditions in dark matter halos are supposedly suitable for the 
existence of a BEC of dark matter particles though, assuming that dark matter is bosonic \cite{2011Mato}. \\
The scenario of a BEC at finite temperatures has never been extended to the example of compact objects, so the present 
work represents the first contribution in this direction. In~\secref{sec:T0} we first review the zero-temperature 
case as presented in Ref. \cite{2011Cha2}, before outlining the contents of the main body of the paper which contains 
our own work in~\secref{sec:presentwork}, including a motivation for the specific choice of treatment.

\subsection{Zero-temperature case}
\label{sec:T0}
A BEC subject to contact and gravitational interaction has been formulated in Ref. \cite{2011Cha2} via 
a Heisenberg equation for the bosonic field operator $\hat{\Psi}(\mb{x},t)$ representing 
bosons with mass $m$. The corresponding second-quantized Hamiltonian operator for this system reads 
\begin{eqnarray} \label{eq:Ham}
  \hat{\mathcal{H}} &=& \int d^3x \,\hat{\Psi}^{\dag}(\mb{x},t) \bigg[ -\frac{\hbar^2}{2m} \Delta - \mu \\ 
  &~& ~~~~~~~~+ \frac{1}{2} \int d^3x' \, \hat{\Psi}^{\dag}(\mb{x}',t) U(\mb{x},\mb{x}') 
    \hat{\Psi}(\mb{x}',t) \bigg] \hat{\Psi}(\mb{x},t) \,, \nonumber
\end{eqnarray}
where $\mu$ denotes the chemical potential in the grand-canonical treatment, and the interaction term 
$U(\mb{x},\mb{x}')$ in the presence of contact and gravitational interaction reads 
\begin{equation} \label{eq:interactions}
  U(\mb{x}-\mb{x}') = g\, \delta(\mb{x}-\mb{x}') - \frac{G m}{|\mb{x}-\mb{x}'|}\,.
\end{equation}
Here $g = 4\pi \hbar^2 a/m$ denotes the strength of the repulsive contact interaction, with $a$ being the 
s-wave scattering length of the bosons in the system, while $G$ is Newton's gravitational constant. The 
resulting Heisenberg equation of motion defined from the Hamiltonian~\eqref{eq:Ham} reads 
\begin{eqnarray} \label{eq:Heisenberg}
  && i\hbar \frac{\partial}{\partial t} \hat{\Psi}(\mb{x},t) = \bigg[ -\frac{\hbar^2}{2m} \Delta \\
    &&~~~~~+ g \big|\hat{\Psi}(\mb{x},t)\big|^2 -\int d^3x' \frac{G m^2}{|\mb{x}-\mb{x}'|} \, 
    \big|\hat{\Psi}(\mb{x},t)\big|^2 \bigg] \hat{\Psi}(\mb{x},t) \,. \nonumber
\end{eqnarray}
To implement the presence of a condensate as well as of thermal and quantum fluctuations, the field operator 
can be split into a mean field condensate and fluctuations. For the zero-temperature case, where no thermal 
fluctuations are present, and weak enough interparticle interactions such that quantum fluctuations can be 
neglected as well, a mean field condensate is assumed, represented by the wave function 
\begin{equation}
  \Psi(\mb{x},t) = \langle \hat{\Psi}(\mb{x},t) \rangle \,.
\end{equation}
The Heisenberg equation~\eqref{eq:Heisenberg} then reduces to the Gross-Pitaevskii (GP) equation, 
\begin{equation} \label{eq:GP}
  i\hbar \frac{\partial}{\partial t} \Psi(\mb{x},t) = \bigg[ -\frac{\hbar}{2m} \Delta
    + g \left|\Psi(\mb{x},t)\right|^2 + \Phi(\mb{x},t) \bigg] \Psi(\mb{x},t) \,,
\end{equation}
where we have defined the Newtonian gravitational potential as 
\begin{equation}
  \Phi(\mb{x},t) = -\int d^3x' \frac{G m^2}{|\mb{x}-\mb{x}'|} \, \big| \Psi(\mb{x}',t)\big|^2 \,.
\end{equation}
Assuming a Madelung representation of the condensate wave function, i.e. using an ansatz 
featuring an amplitude and a phase, 
\begin{equation} \label{eq:ansatz}
	\Psi(\mb{x},t) = \sqrt{n_{\mb{0}}(\mb{x},t)} ~ e^{iS(\mb{x},t)} \,,
\end{equation}
we can identify the density of the condensate as 
\begin{equation}
	n_{\mb{0}}(\mb{x},t) = \left|\Psi(\mb{x},t)\right|^2 \,.
\end{equation}
With~\eqref{eq:ansatz}, the Gross-Pitaevskii equation~\eqref{eq:GP} decomposes into two 
equations by setting its real and imaginary part to zero separately. This results in two coupled 
hydrodynamic equations, i.e. the continuity equation and the Euler equation for the density 
$n_{\mb{0}}$ and for the velocity field $\mb{v}=\hbar \,\nabla S/m$, 
\begin{subequations} 
\begin{align}
 \frac{\partial n_{\mb{0}}}{\partial t} + \nabla\cdot (n_{\mb{0}} \,\mb{v}) &= 0 \,,\\
 m\,n_{\mb{0}} \left[\frac{d\mb{v}}{dt} + (\mb{v}\cdot \nabla) \,\mb{v} \right]  
    &= - \frac{g}{2} \nabla n_{\mb{0}}^2 - m\,n_{\mb{0}} \,\nabla \Phi - 
      \nabla\cdot \sigma^Q_{ij}\,. \label{eq:Euler}
\end{align}
\end{subequations}
The last term in the Euler equation contains the so-called quantum stress tensor 
\begin{equation}
  \sigma^Q_{ij} = \frac{\hbar^2}{4 m}\, n_{\mb{0}} \,\nabla_i \nabla_j \ln n_{\mb{0}} \,,
\end{equation}
which represents a quantum contribution originating from the Laplacian term in the Gross-Pitaevskii equation. 
Commonly the Thomas-Fermi (TF) approximation is adapted, in which the kinetic term is neglected, and the 
quantum stress tensor is dropped. Also all other time dependences are neglected from here on since we restrict 
ourselves to static configurations only. \\
By comparison of~\eqeqref{eq:Euler} with the general form of the Euler equation of a fluid, we can identify 
the pressure of the condensate from the first term on the RHS as 
\begin{equation} 
  p = \frac{g}{2} \,n_{\mb{0}}^2\,.
\end{equation}
It is non-zero even for zero temperature, which is a direct consequence of the presence of the contact 
interaction. For zero contact interaction, the pressure vanishes as well, as should be the case for a free 
Bose gas \cite{1980Land}. Defining the mass density of the system as 
\begin{equation}
  \rho = m\, n_{\mb{0}} \,
\end{equation}
leads to the equation of state 
\begin{equation} \label{eq:EoSwithrho} 
  p = \frac{g}{2m^2} \,\rho^2\,.
\end{equation}
This is a polytropic equation of state, in general written as 
\begin{equation} \label{eq:EoS}
  p = \kappa \, \rho^{\gamma} \,,
\end{equation}
where $\gamma = 1 + 1/n$ defines the polytropic index $n$, and $\kappa$ represents a suitable constant 
of proportionality. In the present case of a BEC  we have $n=1$ and $\kappa = 2\pi \hbar^2 a / m^3$. \\
Neglecting all time dependent terms in~\eqeqref{eq:Euler} and employing the TF approximation leads to 
\begin{equation} \label{eq:Euler2}
 \nabla p = -\rho\, \nabla \Phi\,.
\end{equation}
Combining \eqsref{eq:EoSwithrho},~\eqref{eq:Euler2} and the Poisson equation for the gravitational 
potential, 
\begin{equation} \label{eq:Poisson}
  \nabla^2 \Phi = -4 \pi G \, \rho \,,
\end{equation}
results in the so-called Lane-Emden equation, a second-order differential equation for the mass density of the 
condensate $\rho$ as a function of the radial coordinate $r$. With the substitutions 
$\chi = \left( \rho / \rho_c \right)^{1/n}$, where $\rho_c$ is the central condensate density, as well as 
the dimensionless length scale $\xi = r\, \sqrt{4\pi G / \left[ \kappa (n+1) \rho_c^{-1+1/n} \right]}$, the 
Lane-Emden equation reads 
\begin{equation} 
  \frac{1}{\xi^2} \frac{d}{d\xi}\left( \xi^2 \,\frac{d\chi}{d\xi} \right) = -\chi^n \,.
\end{equation}
For $n=1$ the system can be solved analytically, yielding the corresponding mass limit straightforwardly. 
The exact solution in this case is found as 
\begin{equation} \label{eq:LaneEmden}
  \chi \left(\xi\right) = \frac{\sin \xi}{\xi} \,,
\end{equation}
which gives the radius $R_0$ of the star by the condition $\chi \left(\xi_0 \right) =0$, 
i.e. $\xi_0 = \pi$, yielding the condensate radius 
\begin{equation} \label{eq:R0ChavHark}
   R_0 = \pi \, \sqrt{\frac{\hbar^2 a}{G m^3}} \,.
\end{equation}
The mass of the object can then be obtained by integrating the density profile up to that point, 
\begin{equation} \label{eq:ChavHarkMass} 
  M = 4 \pi^2 \, \left( \frac{\hbar^2 a}{G m^3} \right)^{3/2} \rho_c \,,
\end{equation}
and depends on the condensate density at the center of the star $\rho_c$. These results were already obtained 
in Ref. \cite{2011Cha2} and applied to the example of neutron stars. Some physical criterion has to be invoked 
in order to determine a limit on the maximum mass of the configuration. A limit on the central density can follow 
from demanding that the adiabatic speed of sound in the fluid at the center of the star be bound by the speed of 
light. Alternatively a limiting mass can be calculated from the criterion of gravitational collapse, derived from 
the Schwarzschild radius of the configuration. In Ref. \cite{2011Cha2}, the Schwarzschild limit resulted in a 
maximum mass of about $2.3\, M_{\odot}$. \\
We would like to note that the results for the equation of state can also be used in more general versions of 
the theory, i.e. when extending the treatment to general relativistic settings. Considering the Einstein equations 
with an ansatz for a spherically symmetric metric leads to the  Tolman-Oppenheimer-Volkoff 
equation \cite{1939Tolm,1939Oppe}, 
\begin{equation} \label{eq:TOV}
 \frac{dP(r)}{dr}= -\frac{G\left[ \rho(r) + \frac{P(r)}{c^2} \right]
  \left[ \frac{4\pi P(r) r^3}{c^2} + M(r) \right]}{r^2 \, \left[1-\frac{2G M(r)}{rc^2}\right]}\,.
\end{equation}
This equation, together with an equation of state $p=p(\rho)$ as e.g. given by~\eqref{eq:EoSwithrho}, and the mass 
conservation equation 
\begin{equation} \label{eq:mass}
  \frac{dM(r)}{dr} = 4\pi \rho(r) \,r^2
\end{equation}
completely determines the system in question. In this way, the equation of state extracted from the above 
procedure can be used in the context of general relativity as well. This was worked out for the zero-temperature 
condensate in Ref. \cite{2011Cha2} in addition to the Newtonian case. Alternatively, the equation of state might 
serve as an input parameter in astrophysical simulations for compact objects which do not consider the physics inside 
the star from first principles but approach the issue on a more phenomenological level \cite{2001Latti}.

\subsection{Finite-temperature case applied to neutron stars}
\label{sec:presentwork}
In the work presented in this paper, we carry out a generalization of the above treatment, aiming at 
deriving a theory of a Bose-Einstein condensate subject to repulsive contact interaction and attractive 
gravitational interaction for the case of finite temperatures. A first step in this direction in the framework of the 
Heisenberg equation~\eqref{eq:Heisenberg} was performed in Ref. \cite{2012Hark}, where the field operator is split 
into a mean field contribution 
and a fluctuating term, i.e. $\hat{\Psi}(\mb{x},t) = \langle \hat{\Psi}(\mb{x},t) \rangle + \hat{\psi}(\mb{x},t)$. 
However, the authors solely calculated the equation of state of condensate and thermal density, and applied them 
to the example of dark matter, deriving the resulting expansion behaviour of the universe in a cosmological 
scenario. In our case however, we investigate the behaviour of a self-gravitating Bose-Einstein condensate in compact 
objects, compute the density profiles of a BEC star at finite temperatures and derive relevant macroscopic quantities, 
which can then be compared to astrophysical observations. \\
To do so, we first need to determine the appropriate treatment for the scenario in question. One aspect to be reflected 
upon is the gravitational framework of the theory, i.e. the choice between Newtonian gravity and general relativity. 
Estimating the typical size scales of the system and comparing them to their corresponding Schwarzschild radii, 
\begin{equation}
  r_S = \frac{2G M}{c^2} \,,
\end{equation}
shows whether the general relativistic regime is reached or Newtonian gravity suffices for the description of the 
gravitational interactions. Furthermore, we need to consider the typical velocities of particles in the system in order to 
be able to distinguish between non-relativistic and relativistic dispersion relations. From the typical temperatures 
in compact objects we can estimate the particle velocities from 
\begin{equation}
  v = \sqrt{\frac{2k_B T}{m}} \,,
\end{equation}
and a comparison with the speed of light $c$ will determine the appropriate treatment. For $v \ll c$, we can resort to a 
non-relativistic quantum-mechanical treatment with a Schr\"o\-din\-ger-type equation as outlined above, whereas for 
$v \sim c$, it would be necessary to formulate the theory in terms of a relativistic description with the 
Klein-Gordon equation. \\
The case of a neutron star can at least partly be treated with a non-relativistic dispersion relation, since typical 
temperatures range from $10^{11} - 10^{12} \, \mathrm{K}$ at the initial stages, and decrease down to $10^6 \,\mathrm{K}$ after 
several years, corresponding to thermal velocities of $0.09 \, c$ and $0.3\cdot 10^{-3}\, c$, respectively. 
As for the gravitational theory, the typical size of a neutron star is estimated to be about $12\, \mathrm{km}$, and at the 
observed masses between $1 - 2 \,M_{\odot}$, typical radii are only about $2 - 4$ times larger than the respective 
Schwarzschild radii, which means that a general relativistic description should be necessary. \\
Despite these numbers, for the sake of simplicity here we develop a theory which is non-relativistic in both regards, i.e. 
a model for a non-relativistic BEC in Newtonian gravity, and evaluate later to what extent the theory is applicable to neutron 
stars. We treat the system in the framework of a Hartree-Fock theory, and set up self-consistency equations for 
the densities of the BEC and the thermal cloud of excited atoms. To this end we start from a general 
Hamiltonian and derive the governing Hartree-Fock equations for the wave functions of the particles in the ground state 
and in the thermally excited states. The detailed derivations of this part are shown in the appendix, as the Hartree-Fock 
theory for bosons has been worked out in the literature before, see e.g. Ref. \cite{1997Oehb}. Still for the general case of 
a Hamiltonian with unspecified interactions $U(\mb{x},\mb{x}')$ we then consider the semi-classical limit of the 
theory and derive the equations for the macroscopic densities of condensate and thermal excitations. In~\secref{sec:CG}, 
we start from the respective equations of motion in the semi-classical approximation for the case of contact and gravitational 
interaction. We show the numerical solution of the system of equations in~\secref{sec:Num}, and then derive astrophysical 
consequences and quantities in~\secref{sec:astro}, like the size scales and maximum mass of the system and the equation of 
state of matter inside the star. We investigate the physical viability 
of the system and obtain a limit for the maximally possible masses in analogy to the TOV-limit. In~\secref{sec:Concl} 
ultimately, we comment on the significance of our work in the astrophysical context and conclude the part with an 
outlook to further investigations. \\ ~\\ ~\\

\section{Semi-classical Hartree-Fock theory for contact and gravitational interaction}
\label{sec:CG}
In this section, we first revisit the Hartree-Fock equations of motion governing the evolution of the condensate and thermal 
density in the semi-classical approximation as derived in detail in the appendix. Then we show how to solve the combined system of 
self-consistency equations in two regimes, distinguished by the presence and vanishing of the condensate, respectively. 
Originating from the Hamiltonian~\eqref{eq:Ham} of the system with the interactions~\eqref{eq:interactions}, 
a Hartree-Fock theory was developed, resulting in the equations of motion for the wave functions of condensate 
and thermal fluctuations calculated from a variation of the free energy with respect to one-particle wave 
function basis of the system. After having obtained the exact self-consistency equations governing the system, 
the semi-classical limit of the theory was taken. The detailed derivations are to be found in the appendix.

\subsection{Semi-classical equations of motion}
In this section we consider the semi-classical Hartree-Fock equations of motion as derived in the appendix for a system 
with contact and gravitational interaction. Note that we are employing the Hartree approximation for the gravitational 
part of the interactions, i.e. we discard any bilocal contribution to the equations. The equation of motion for the 
condensate density~\eqeqref{eq:HF1CG} and the thermal energies $\epsilon_{\mb{k}}(\mb{x})$ given by~\eqeqref{eq:ekCG} 
thus read 
\begin{eqnarray} \label{eq:H1/2F}
    && - \mu + g\,\left[ n_{\mb{0}}(\mb{x}) + 2\, n_{\mathrm{th}}(\mb{x}) \right] \\
    && ~~~~~~~~~~~~~~~~ - \int d^3x' \, \frac{G m^2}{|\mb{x}-\mb{x}'|} 
      \, \Bigg[ n_{\mb{0}}(\mb{x}') + n_{\mathrm{th}}(\mb{x}') \Bigg] =0 \,, \nonumber
\end{eqnarray}
and
\begin{eqnarray} \label{eq:ekH1/2F}
    && \epsilon_{\mb{k}}(\mb{x}) = \frac{\hbar^2 \mb{k}^2}{2m} + 2 g\, \big[ n_{\mb{0}}(\mb{x}) 
	+ n_{\mathrm{th}}(\mb{x}) \big] \\
    && ~~~~~~~~~~~~~~~~~~~- \int d^3x' \, \frac{G m^2}{|\mb{x}-\mb{x}'|} \, \bigg[n_{\mb{0}}(\mb{x}') 
	+ n_{\mathrm{th}}(\mb{x}') \bigg] \,. \nonumber
\end{eqnarray}
The first equation is valid for a non-vanishing condensate density, and originates from an equation with the complementary 
solution $n_{\mb{0}}(\mb{x})=0$, as argued already in the appendix. The second Hartree-Fock equation in the semi-classical 
approximation yields the wave vector dependence of the thermal energies, which can be employed to calculate the thermal density from 
its semi-classical definition~\eqref{eq:nthSC} according to 
\begin{equation} \label{eq:nthintermed}
    n_{\mathrm{th}}(\mb{x})= \int \frac{d^3k}{(2\pi)^3} \,\frac{1}
    {e^{\beta \left[ \epsilon_{\mb{k}}(\mb{x}) - 
    \mu \right]} -1} \,.
\end{equation}
In the following, we substitute $\epsilon = \hbar^2 \mb{k}^2/(2m)$, and introduce the abbreviation 
\begin{equation}
  \alpha(\mb{x}) = 2 g\, \big[ n_{\mb{0}}(\mb{x}) + n_{\mathrm{th}}(\mb{x}) \big] + \Phi(\mb{x}) - \mu \,,
\end{equation} 
with the gravitational potential $\Phi(\mb{x})$ now defined as 
\begin{equation} \label{eq:GravPot}
  \Phi(\mb{x}) = - \int d^3x' \, \frac{G m^2}{|\mb{x}-\mb{x}'|} \, \bigg[n_{\mb{0}}(\mb{x}') + 
    n_{\mathrm{th}}(\mb{x}') \bigg] \,.
\end{equation} 
With this, the thermal density~\eqref{eq:nthintermed} becomes 
\begin{equation} 
    n_{\mathrm{th}}(\mb{x})= \frac{\sqrt{2}}{(2\pi)^3} \frac{m^{3/2}}{\hbar^3} 
      \int_0^{\infty} d\epsilon\, \sqrt{\epsilon} \, \frac{1}{e^{\beta \left( \epsilon + \alpha \right)} -1} \,,
\end{equation}
which can be solved with the help of a standard integral \cite{1965Grad}, and yields 
\begin{equation} \label{eq:nthint2}
  n_{\mathrm{th}}(\mb{x}) = \frac{1}{\lambda^3} \, \zeta_{3/2} \left( e^{-\beta \alpha(\mb{x})} \right) \,,
\end{equation}
where $\lambda = (2\pi \beta \hbar^2/m)^{1/2}$ denotes the thermal de Broglie wavelength, and 
\begin{equation}
  \zeta_{\nu} (z) = \sum\limits_{m=1}^{\infty} \frac{z^m}{m^{\nu}} 
\end{equation}
represents the polylogarithmic function.

\subsection{Introduction of spherical coordinates}
Before we proceed to process the derived expressions, we simplify the equations by 
assuming spherical symmetry which enables us to introduce spherical coordinates. Thus, 
both condensate and thermal density simplify to 
\begin{equation}
  n_{\mb{0}}(\mb{x}) = n_{\mb{0}}(r) \,,\quad n_{\mathrm{th}}(\mb{x}) = n_{\mathrm{th}}(r) \,.
\end{equation}
Furthermore, we reformulate the gravitational potential~\eqref{eq:GravPot} in terms of a 
multipole expansion in spherical coordinates. Separating the areas of $r\leq r'$ and 
$r\geq r'$, we express the $1/r$-term in the gravitational potential~\eqref{eq:GravPot} 
as 
\begin{eqnarray} \label{eq:multexp}
 && \frac{1}{|\mb{x}-\mb{x}'|}  = 
    \sum_{l=0}^{\infty} \,\sum_{m=-l}^l \, \frac{4\pi}{2l+1} \, Y_{lm} (\Omega)\, Y^*_{lm} (\Omega') \\
 && ~~~~~~~~~~~~~~~~~~ \times \left[ \Theta(r-r') \,\frac{r'^{l+2}}{r^{l+1}}  + 
    \Theta(r'-r) \,\frac{r^l}{r'^{l-1}} \right] \,.\nonumber
\end{eqnarray}
Applying these substitutions to the Hartree-Fock equations~\eqref{eq:H1/2F} and~\eqref{eq:nthint2}, we 
use the mathematical properties of the spherical harmonics, like the addition theorem, 
\begin{equation} \label{eq:addtheorem}
  \sum_{m=-l}^{l} Y^*_{lm} (\Omega) \,Y^{~}_{lm} (\Omega') = \frac{2l+1}{4\pi} \,,
\end{equation}
the normalization condition, 
\begin{equation}
  \int d\Omega \, Y^*_{lm} (\Omega) \,Y^{~}_{l'm'} (\Omega) = \delta_{ll'} \delta_{mm'} \,,
\end{equation}
and the fact that $Y_{00}(\Omega) = 1/ \sqrt{4\pi}$. With this, the first Hartree-Fock 
equation~\eqref{eq:H1/2F} yields 
\begin{equation} \label{eq:HF1spher}
  - \mu + g\, \Big[ n_{\mb{0}}(r) + 2 n_{\mathrm{th}}(r) \Big] + \Phi(r) =0 \,,
\end{equation}
where the gravitational potential~\eqref{eq:GravPot} now reads in spherical coordinates, 
\begin{eqnarray} \label{eq:Phiofr}
  \Phi(r) &=& - 4\pi G m^2 \, \Bigg\{ \frac{1}{r}  \, \int_0^r dr' \, r'^2 \Big[ n_{\mb{0}}(r') +  n_{\mathrm{th}}(r') \Big] \\
    &~&~~~~~~~~~~~~~~~~~ + \int_r^{\infty} dr' \, r' \Big[ n_{\mb{0}}(r') +  n_{\mathrm{th}}(r') \Big] \Bigg\} \,.\nonumber
\end{eqnarray}
The thermal density~\eqref{eq:nthint2} correspondingly becomes 
\begin{equation} \label{eq:nthexact} 
  n_{\mathrm{th}}(r) = \frac{1}{\lambda^3} \,\zeta_{3/2} \bigg[ e^{-\beta \big( 2g \,
    \left[n_{\mb{0}}(r)+n_{\mathrm{th}}(r)\right] + \Phi (r) -\mu \big)} \bigg] \,.
\end{equation}
Note that this result for the thermal density is valid everywhere in the system. The argument of 
the exponent contains an expression which depends on the radial coordinate. For our system, we expect 
two regimes: the inner zone, where the condensate density is nonzero and coexists with the thermal 
density, and the outer regime, where the condensate vanishes, but a thermal phase continues 
to exist. The boundary between those two regions is given by the Thomas-Fermi radius, i.e. 
the point where the condensate density vanishes, 
\begin{equation} \label{eq:n0R0}
  n_{\mb{0}}(R_0) = 0 \,.
\end{equation}
Therefore, we have to consider two different versions of the thermal density for the inner 
and outer regime, which will be denoted by subscripts $1$ and $2$, respectively. The condensate 
exists solely in the inner region, and is zero outside the Thomas-Fermi radius. \\
In the following two subsections, we will treat both regimes in more detail and further process 
the equations for the condensate and the thermal densities analytically up to a point, where we then 
have to resort to numerical solution methods.

\subsection{Inner regime}
\label{sec:inner}
In the inner regime, we can employ the first Hartree-Fock equation~\eqref{eq:HF1spher} to 
simplify the argument of the exponent in the thermal density~\eqref{eq:nthexact} and obtain 
\begin{equation} \label{eq:nth1}
  n_{\mathrm{th},1}(r) = \frac{1}{\lambda^3} \,\zeta_{3/2} \Big[ e^{- \beta g \, n_{\mb{0}}(r) } \Big] \,.
\end{equation}
Having obtained this expression for the thermal density in the inner regime, 
we can now consider the first Hartree-Fock equation~\eqref{eq:HF1spher}, 
\begin{equation} \label{eq:HF1TF}
  - \mu + g \, n_{\mb{0}}(r) + 2 g \, n_{\mathrm{th},1}(r) + \Phi(r) = 0 \,,
\end{equation}
in order to obtain a solution for the condensate density, and subsequently calculate the thermal 
density in the inner region via~\eqref{eq:nth1}. The first Hartree-Fock equation~\eqref{eq:HF1TF} 
can be further processed by multiplying the equation by $r$ and differentiating twice with respect 
to $r$ to get rid of the integrals which are due to the gravitational interactions. With this, the 
integral equation~\eqref{eq:HF1TF} reduces to a differential equation 
\begin{equation} \label{eq:HF1diffOrig} 
	\frac{\partial^2}{\partial r^2} \bigg\{ r \, \Big[ n_{\mb{0}}(r) + 2 n_{\mathrm{th},1}(r) \Big] \bigg\} = 
		- \sigma^2 \, r \, \bigg[ n_{\mb{0}}(r) + n_{\mathrm{th},1}(r) \bigg] \,,
\end{equation}
where we introduced the inverse length scale 
\begin{equation} \label{eq:sigma}
  \sigma = \sqrt{ \frac{4\pi \, G m^2}{g} } \,,
\end{equation}
which characterizes the typical size scales of the system. Employing~\eqref{eq:nth1}, we can 
express~\eqeqref{eq:HF1diffOrig} only in terms of the condensate density, 
\begin{eqnarray} \label{eq:HF1diff} 
  && \frac{\partial^2}{\partial r^2} \bigg\{ r \, n_{\mb{0}}(r) + 
    \frac{2r}{\lambda^3} \,\zeta_{3/2} \Big[ e^{- \beta g \, n_{\mb{0}}(r) } \Big] \bigg\} \\
  &&~~~~~~~~~~~~ = - \sigma^2 \, r \, \bigg\{ n_{\mb{0}}(r) + 
    \frac{1}{\lambda^3} \,\zeta_{3/2} \Big[ e^{- \beta g \, n_{\mb{0}}(r) } \Big] \bigg\} \,.\nonumber
\end{eqnarray}
This second-order differential equation for $n_{\mb{0}}(r)$ has to be solved by taking into account the 
boundary conditions 
\begin{equation} \label{eq:HF1bound}
  n_{\mb{0}}(0) = A\,, \quad \frac{d n_{\mb{0}}}{dr} \bigg|_{r=0} = 0 \,.
\end{equation}
Here, the constant $A$ represents a parameter which is indirectly related to the total number of particles $N$ of 
the system. It is the only parameter needed in the complete solution of the system in both regimes, and thus the 
choice of $A$ is equivalent to a choice of $N$. After having obtained the numerical solution for the condensate 
density, the thermal density can then be obtained from the result for $n_{\mb{0}}(r)$ using~\eqeqref{eq:nth1}. \\
In the limit of zero temperature, the thermal fluctuations are zero, and from~\eqref{eq:HF1diff} follows that the 
condensate density is exactly determined from the simplified differential equation 
\begin{equation} \label{eq:HF1simple} 
	\frac{\partial^2}{\partial r^2} \bigg[ r \, n_{\mb{0}}(r) \bigg] = 
		- \sigma^2 \, r \, n_{\mb{0}}(r) \,.
\end{equation}
The solution of~\eqref{eq:HF1simple} with~\eqref{eq:HF1bound} is 
\begin{equation} \label{eq:n0zeroT}
	n_{\mb{0}}(r) = A\, \frac{\sin (\sigma r)}{r} \,,
\end{equation} 
which corresponds to the solution~\eqref{eq:LaneEmden} outlined in~\secref{sec:T0}. In this special case, the 
integration constant $A$ can be determined analytically by computing the total number of particles in the system, 
\begin{equation}
  N = 4\pi \int_0^{R_0} dr\, r^2 \, n_{\mb{0}}(r) \,,
\end{equation}
yielding 
\begin{equation}
   A (T=0) = \frac{N}{4\pi^2} \,.
\end{equation}
For zero temperature, it is also possible to calculate the Thomas-Fermi radius $R_0$ according to~\eqref{eq:n0R0}, 
yielding 
\begin{equation} \label{eq:TFT0analytic}
  R_0 = \frac{\pi}{\sigma} \,,
\end{equation}
which coincides with~\eqref{eq:R0ChavHark} due to~\eqeqref{eq:sigma}. For non-zero temperatures, the Thomas-Fermi 
radius will differ from this value, since the condensate density obtains corrections due to thermal fluctuations.

\subsection{Outer regime}
In the outer regime, the thermal density~\eqref{eq:nthexact} is specified further by considering 
the fact that $n_{\mb{0}}(r)=0$. The thermal density then reads 
\begin{equation} \label{eq:nth2}
  n_{\mathrm{th},2}(r) = \frac{1}{\lambda^3} \,\zeta_{3/2} \bigg[ e^{-\beta \big( 
  2g \,n_{\mathrm{th},2}(r) + \Phi (r) -\mu \big)} \bigg] \,,
\end{equation}
where the gravitational potential~\eqref{eq:Phiofr} is evaluated for $r>R_0$ as 
\begin{eqnarray} \label{eq:Phi2orig}
  && \Phi(r) = - 4 \pi G m^2 \, \Bigg\{ \frac{1}{r}  \, \int_0^{R_0} dr' \, r'^2 
    \Big[ n_{\mb{0}}(r') +  n_{\mathrm{th},1}(r') \Big] \nonumber\\
  &&~~ + \frac{1}{r}  \, \int_{R_0}^{r} dr' \, r'^2 
    n_{\mathrm{th},2}(r') + \int_{r}^{\infty} dr' \, r' n_{\mathrm{th},2}(r') \Bigg\} \,.
\end{eqnarray}
Note that $\Phi(r)$ still contains the condensate density in the first term, since the 
presence of the condensate in the inner regime gravitationally influences the thermal density 
in the outer region. However, this dependence can be simplified in notation by introducing the 
number of condensed atoms, 
\begin{equation} \label{eq:numbers1}
  N_{\mb{0}} = 4\pi \,\int_0^{R_0} dr \, r^2 \, n_{\mb{0}}(r) \,,
\end{equation}
and the number of thermal atoms in the inner regime, 
\begin{equation} \label{eq:numbers2}
  N_{\mathrm{th},1} = 4 \pi \int_0^{R_0} dr \, r^2 \, n_{\mathrm{th},1}(r) \,.
\end{equation}
For abbreviation, we denote the total number of particles in the inner regime as 
\begin{equation} \label{eq:Nin}
  N_{\mathrm{in}} = N_{\mb{0}} + N_{\mathrm{th},1} \,.
\end{equation}
The gravitational potential in the outer region~\eqref{eq:Phi2orig} then simplifies to 
\begin{eqnarray} \label{eq:Phi2}
  && \Phi(r) = - \frac{G m^2 N_{\mathrm{in}} }{r} \\
  && - 4 \pi G m^2 \, \Bigg[ 
    \frac{1}{r}  \, \int_{R_0}^{r} dr' \, r'^2 
    n_{\mathrm{th},2}(r') + \int_{r}^{\infty} dr' \, r' n_{\mathrm{th},2}(r') \Bigg] \,.\nonumber
\end{eqnarray}
The determining equation~\eqref{eq:nth2} for $n_{\mathrm{th},2}(r)$ is rather involved due 
to the polylogarithmic function and the occurrence of the thermal density as the argument of the integral in the 
gravitational potential~\eqref{eq:Phi2}. In order to solve the equation, we will carry out some substitutions to 
convert the integral equation to a differential one. First, we integrate expression~\eqref{eq:nth2} over the 
region outside of the Thomas-Fermi radius, i.e. over the regime $r\in [R_0,\infty]$. Substituting this integral 
with a function $h(r)$, defined by 
\begin{equation} \label{eq:hofr}
 h(r) := \frac{1}{r}  \, \int_{R_0}^r dr' \, r'^2 \, n_{\mathrm{th},2}(r') 
    + \int_r^{\infty} dr' \, r'\,  n_{\mathrm{th},2}(r') \,,
\end{equation}
Equation~\eqref{eq:nth2} then reads 
\begin{eqnarray} \label{eq:nthout2}
  h(r) &=& \frac{1}{\lambda^3} \bigg\{ \frac{1}{r}  \, \int_{R_0}^r dr' \, r'^2 \, 
    \zeta_{3/2} \left[ z(r') \right] \\
  &~& ~~~~~~~~~~+ \int_r^{\infty} dr' \, r'\, \zeta_{3/2} \left[ z(r') 
    \right)] \bigg\} \,, \nonumber
\end{eqnarray}
with the argument 
\begin{eqnarray}
  && z(r) = \mathrm{Exp} \Bigg\{ -\beta \bigg[ -\frac{2g}{r} \, \frac{d^2}{d r^2} \left[ r\, h(r) \right] \\
  &&~~~~~~~~~~~~~~~ - \frac{G m^2 N_{\mathrm{in}} }{r} - 4\pi G m^2\, h(r) - \mu \bigg] \Bigg\} \,. \nonumber
\end{eqnarray}
The thermal density can be obtained by multiplying $h(r)$ with $r$ and differentiating twice, i.e. 
\begin{equation} \label{eq:nth2fromh}
  n_{\mathrm{th},2}(r) = -\frac{1}{r} \, \frac{d^2}{dr^2} \bigg[ r\, h(r) \bigg] \,.
\end{equation}
We also have to insert an expression for the chemical potential into the 
equation. It is obtained by evaluating the first Hartree-Fock equation~\eqref{eq:HF1TF} at the 
Thomas-Fermi radius $r=R_0$ as 
\begin{equation}
  \mu = 2 g\, n_{\mathrm{th},1}(R_0) + \Phi(R_0) \,,
\end{equation}
which yields with~\eqref{eq:nth1},~\eqref{eq:Phi2} and~\eqref{eq:hofr}
\begin{equation}
  \mu = \frac{2g}{\lambda^3}\, \zeta_{3/2}(1) - \frac{G m^2 N_{\mathrm{in}} }{R_0} 
  - 4 \pi G m^2 \,h(R_0) \,. 
\end{equation}
By multiplying~\eqeqref{eq:nthout2} with $r$ and differentiating twice with respect to $r$ 
we end up with a differential equation for $h(r)$, 
\begin{equation} 
  \frac{d^2}{d r^2} \bigg[r\, h(r) \bigg] = 
    - \frac{r}{\lambda^3} \, \zeta_{3/2}\left[ z(r) \right] \,,
\end{equation} 
with the argument 
\begin{eqnarray}
  && z(r) = \mathrm{Exp} \bigg( -\beta \bigg\{ -\frac{2g}{r} \,
    \frac{d^2}{d r^2} \left[ r\, h(r) \right] - \frac{2g}{\lambda^3}\, \zeta_{3/2}(1) \\
  &&~~~~~  - Gm^2 \left[ 4\pi \left[ h(r) - h(R_0) \right]
    - N_{\mathrm{in}} \left( \frac{1}{r} - \frac{1}{R_0} 
    \right) \right] \bigg\} \bigg) \,. \nonumber
\end{eqnarray}
For convenience we will carry out another substitution, i.e. 
\begin{equation} \label{eq:Hr}
  H(r) = h(r) - h(R_0) \,.
\end{equation}
This eliminates the unknown $h(R_0)$-term in the exponent, while~\eqref{eq:nth2fromh} is conserved 
in its form, 
\begin{equation} \label{eq:hnth}
  n_{\mathrm{th},2}(r) = -\frac{1}{r} \, \frac{d^2}{dr^2} \bigg[ r\, H(r) \bigg]\,.
\end{equation}
The final differential equation for $H(r)$ thus reads 
\begin{eqnarray} \label{eq:nthoutFinal}
  && \frac{d^2}{d r^2} \bigg[r\, H(r) \bigg] = \\
  && ~~~~~~ - \frac{r}{\lambda^3} \, \zeta_{3/2}\bigg(  
    \mathrm{Exp} \bigg\{ -\beta \bigg[ -\frac{2g}{r} \, 
    \frac{d^2}{d r^2} \left[ r\, H(r) \right] \nonumber \\
  && ~~~~~~ - 4\pi G m^2 H(r) - \frac{2g}{\lambda^3}\, \zeta_{3/2}(1) \nonumber \\
  && ~~~~~~ - G m^2 N_{\mathrm{in}} \left( \frac{1}{r} - \frac{1}{R_0} 
    \right) \bigg] \bigg\} \bigg) \,. \nonumber
\end{eqnarray} 
In order to solve it in the outer regime for $r>R_0$, we have to specify appropriate boundary conditions. 
From the definition of $H(r)$ in~\eqref{eq:Hr}, we deduce the condition 
\begin{equation} \label{eq:bc1}
  H(R_0) = 0\,.
\end{equation}
Furthermore, we have to demand that the thermal densities of inner and outer regime must be equal at 
the Thomas-Fermi radius, i.e. 
\begin{equation}  
  n_{\mathrm{th},1}(R_0) = n_{\mathrm{th},2}(R_0) \,.
\end{equation}
From the relation~\eqref{eq:hnth} between $n_{\mathrm{th},2}(r)$ and $H(r)$ as well as~\eqref{eq:nth1}, we end 
up with the second boundary condition 
\begin{equation} \label{eq:bc2}
  \frac{1}{\lambda^3} \zeta_{3/2}(1) = -H''(R_0) - \frac{2}{R_0} H'(R_0) \,.
\end{equation}
Solving~\eqref{eq:nthoutFinal} with the boundary conditions~\eqref{eq:bc1} and~\eqref{eq:bc2} 
thus determines the thermal density via~\eqref{eq:hnth} in the outer region.

\section{Numerical Simulations and Solution}
\label{sec:Num}
We will now proceed with describing the numerical procedure to solve the coupled equations 
for the two densities as outlined in the previous section. We distinguish two regimes, the condensate 
area, $0\leq r \leq R_0$, and the outer area, $r > R_0$, where the condensate density $n_{\mb{0}}(r)$ 
vanishes. The thermal density $n_{\mathrm{th}}(r)$ is nonzero in both regimes. We have to solve the 
equation~\eqref{eq:HF1diff} for the condensate density in the inner regime using the boundary 
conditions~\eqref{eq:HF1bound}, which will further determine the thermal density in the inner regime 
via~\eqref{eq:nth1}; whereas for the outer regime we have to solve~\eqeqref{eq:nthoutFinal} with the 
boundary conditions~\eqref{eq:bc1} and~\eqref{eq:bc2} to obtain the thermal density in the outer regime 
via~\eqref{eq:hnth}. 
Note that in the whole procedure we do not need to specify the chemical potential $\mu$ since we have 
managed to eliminate or substitute it wherever it occurred. Instead, however, the constant $A$ appears 
in~\eqref{eq:HF1bound}, as a yet unknown parameter connected to the total number of particles. The correct 
value of A can only be 
determined numerically after having obtained the solution, i.e. in order to carry out the simulation for 
a fixed total number of particles, the parameter $A$ has to be tuned to achieve a specific $N$. Important 
to note is the fact that $A$ is the only input parameter to our solution, to be specified for the interior 
regime. For the solution in the outer regime, results from the inner region are used as parameters, 
i.e. the number of particles~\eqref{eq:Nin} as well as the Thomas-Fermi radius $R_0$ from~\eqref{eq:n0R0}. 
Apart from these values, however, no additional parameters are necessary in the outer regime, and thus 
the complete solution of the system in both the inner and outer region is determined only by specifying 
the parameter $A$. 

\subsection{Dimensionless parameters}
In order to carry out the numerical calculations cleanly, we rewrite all expressions using dimensionless 
quantities according to $r \rightarrow \rho = \sigma r$, $T \rightarrow \theta = T/T_{\mathrm{ch}}$, 
$n \rightarrow \tilde{n} = n \, \lambda_{\mathrm{ch}}^3 = n\, g/k_B T_{\mathrm{ch}}$ and 
$\epsilon \rightarrow \tilde{\epsilon} = \epsilon/k_B T_{\mathrm{ch}}$, 
where $n$ stands for a particle number density and $\epsilon$ for an energy. Any other quantity,  
when expressed with a tilde, as e.g. $\tilde{\mu}$ or $\tilde{\Phi}$, denotes the corresponding 
dimensionless quantity. The newly introduced constants are a characteristic temperature for the 
system in question, and the corresponding de Broglie wavelength, 
\begin{equation}
	T_{\mathrm{ch}} = \frac{\hbar^2 \pi}{2 a^2 m k_B} \,,\quad
	\lambda_{\mathrm{ch}} = \sqrt{\frac{2\pi \hbar^2}{m k_B T_{\mathrm{ch}}}} = 2 a \,.
\end{equation} 
The inverse length scale $\sigma$ has been introduced before in~\eqeqref{eq:sigma} and determines 
the typical size scale of the system in question. In the following, we will elaborate on the concrete 
values of all parameters used in the computations.

\subsection{Simulation details and results} 
\label{sec:simdet}
In order to calculate a solution to the above equations, we have to decide upon a specific 
application of our theory. Choosing the case of a superfluid neutron star with neutron pairs as 
the effective bosons in the system, we have to adjust the simulation parameters to the conditions  
within these objects. We will resort to observational information to fix the appropriate range of 
parameters in order to be in accordance with physically realistic scenarios. \\
Considering the typical masses of neutron stars and the mass of a neutron pair, we carried out 
the simulation for a total number of particles of $N_{\mathrm{tot}}= 10^{57}$, which results in a neutron 
star of about $1.7 \, M_{\odot}$. The parameter that controls the total number of particles in the 
Hartree-Fock theory is $A$. Thus we have to tune the value of $A$ in order to obtain such a specific 
number of particles. \\
A microscopic parameter to be determined is the contact interaction strength $g$, which in 
turn depends on the s-wave scattering length $a$ of the neutron pairs inside the star. As a rough 
estimate for $a$ within the hard sphere scattering approximation, we will use the average volume 
which is to be expected for each particle in the star. With typical radii of neutron stars of about 
$12 \,\mathrm{km}$, and a total number of particles of $10^{57}$, each particle can move within a spherical 
volume of radius $10^{-15} \,\mathrm{m}$, so we choose $a=1\,\mathrm{fm}$. \\
Temperatures in a neutron star depend on its stage of evolution, and range from $10^{12} \,\mathrm{K}$ 
at the time of its formation down to $10^6 \,\mathrm{K}$ after a rapid cooling stage of several years. Thus, 
there is a broad spectrum of temperatures possible. In our simulations, we used a range of temperatures 
between $10^{11} \,\mathrm{K}$ and $4\cdot 10^{11}\, \mathrm{K}$, which cover the high end of the possible 
temperature regime. The reason for choosing such high temperatures lies in the results themselves: we 
found the thermal fluctuations negligible for temperatures below $10^{11}\, \mathrm{K}$, implying that in 
that range the zero-temperature treatment would be sufficient. On the other hand, numerical computations 
for higher temperatures than $4\cdot 10^{11}\, \mathrm{K}$ become unstable, thus providing a natural upper 
limit of our investigations. \\
\begin{figure}
	\centering
	\includegraphics[width=0.4\textwidth]{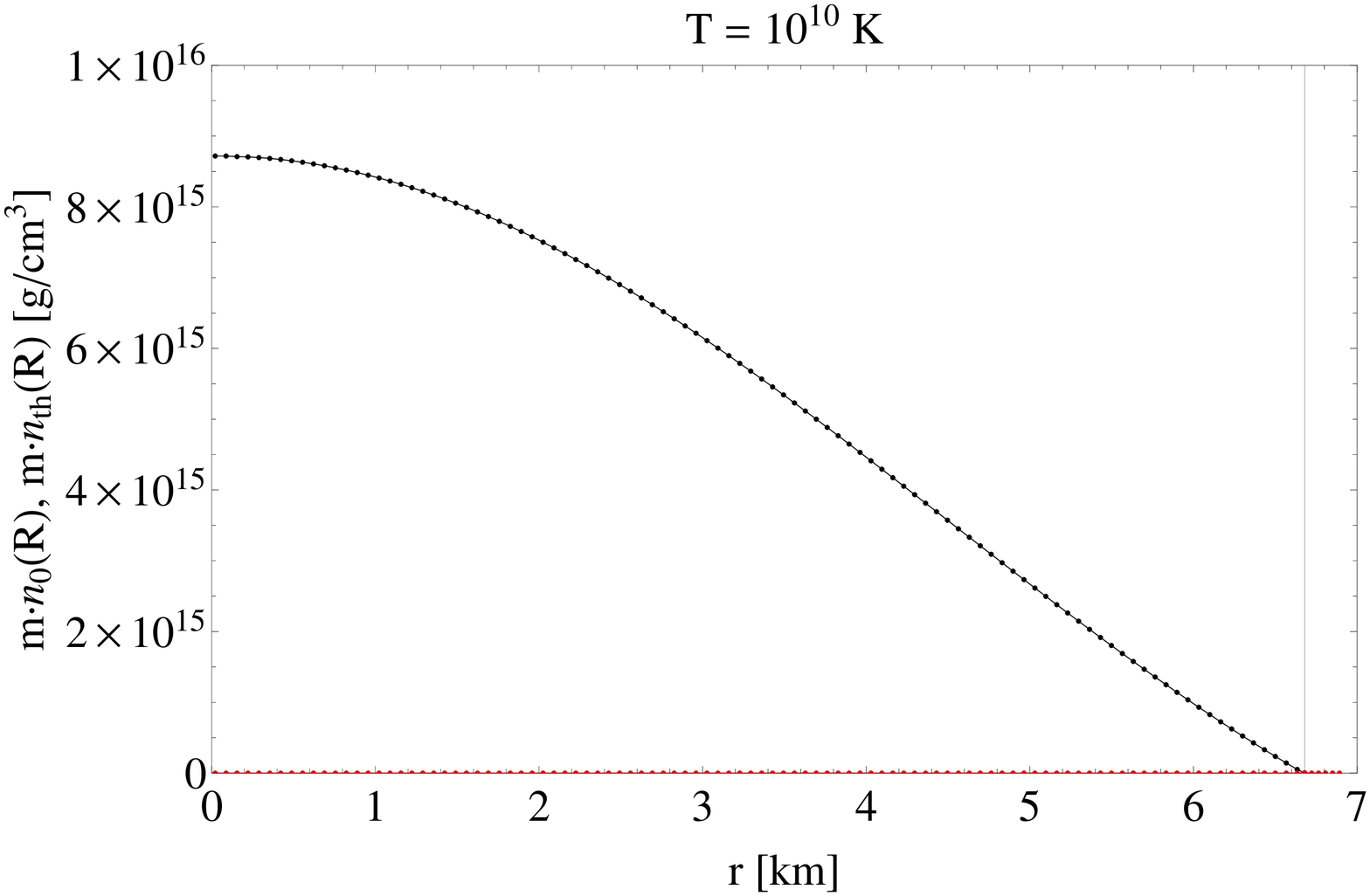}
	\includegraphics[width=0.4\textwidth]{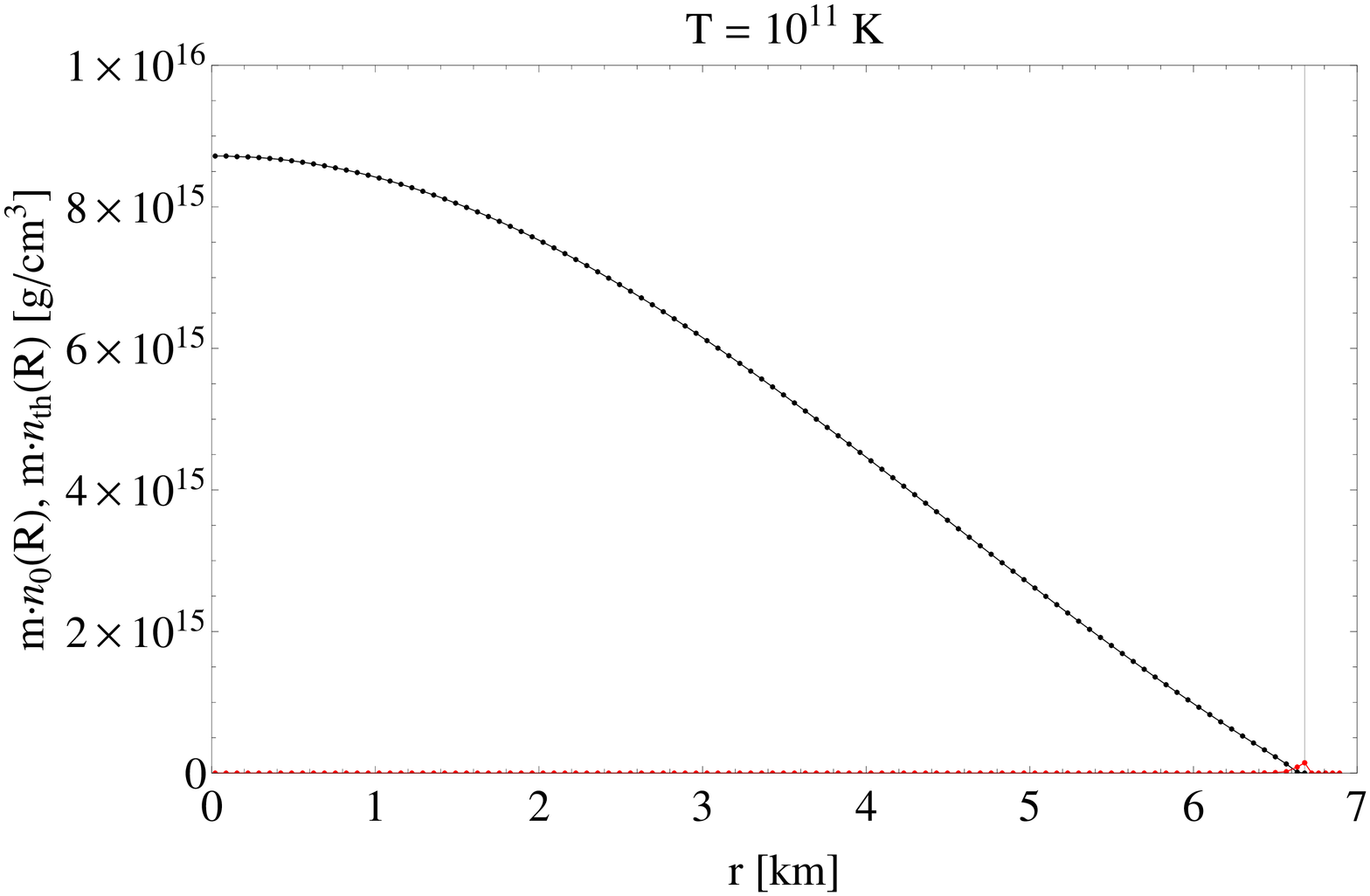}\\
	\includegraphics[width=0.4\textwidth]{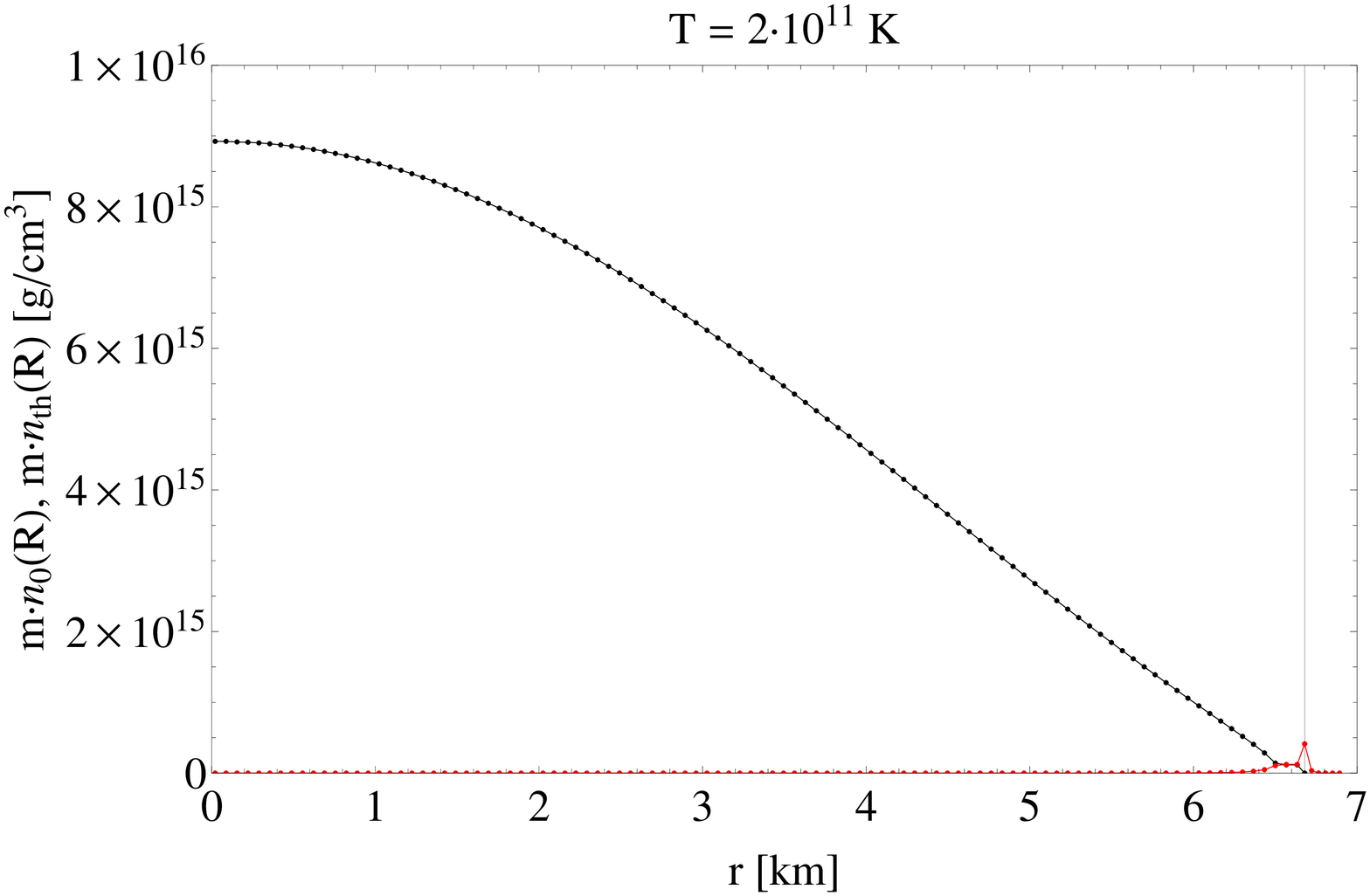}
	\includegraphics[width=0.4\textwidth]{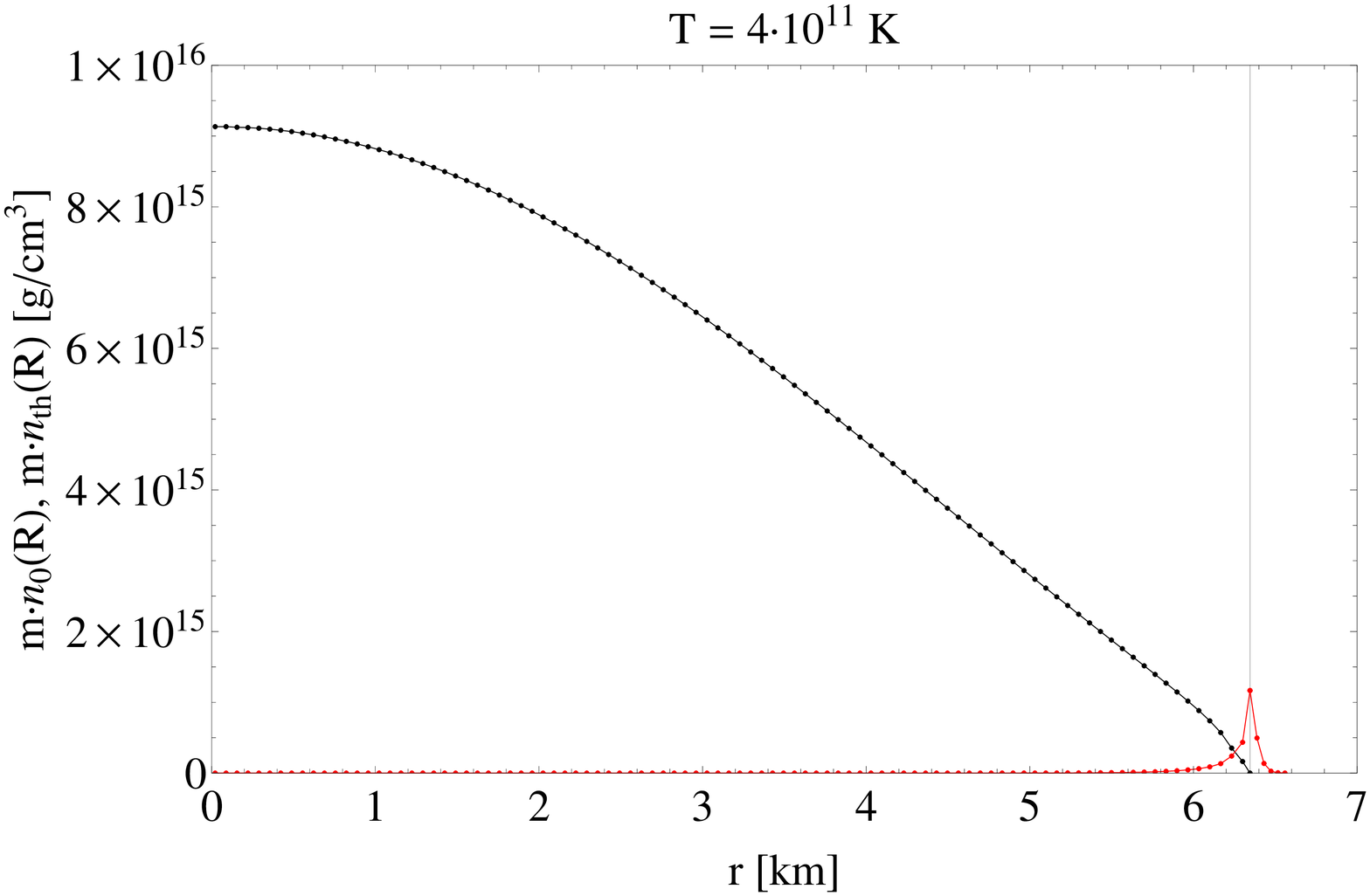}
	\caption{Radial profiles of condensate (black) and thermal density (red) 
	with increasing temperature for the total particle number $N_{\mathrm{tot}} = 10^{57}$.}
	\label{fig:densityprofiles}
\end{figure}
\begin{table*}
    \centering
    \begin{tabular}{c|c|c|c|c|c}
    \hline \hline 

    \hline 
    T [K]  & $0$ & $10^{11}$ & $2\cdot 10^{11}$ & $4\cdot 10^{11}$ & $6\cdot 10^{11}$ \\
    \hline \hline
    $\rho_c$ [$10^{15} \mathrm{g}/ \mathrm{cm}^3$]  & $8.718$ & $8.718$ & $8.925$ & $9.133$ & $9.341$ \\ 
    \hline
    $R_0$ [km] & $~~6.701~~$ &  $6.680$ & $6.546$ & $6.346$ & $6.012$ \\ 
    \hline
    $R_{\mathrm{th}}$ [km] & -- &  $~~6.936~~$ & $~~6.802~~$ & $~~6.602~~$ & $~~6.268~~$ \\ 
    \hline

    \hline \hline
    \end{tabular}
\caption{Summary of simulated data: central condensate density $\rho_c$, Thomas-Fermi radius 
$R_0$, and total radius of the star $R_{\mathrm{th}}$. }
\label{tab:summary}
\end{table*}
For the outlined values of the parameters, the inverse length scale $\sigma$ is computed 
from \eqref{eq:sigma} as 
\begin{equation}
  \sigma \simeq 4.69 \cdot 10^{-4} \,\mathrm{m}^{-1} \,,
\end{equation}
which leads to a Thomas-Fermi radius at zero temperature~\eqref{eq:TFT0analytic} of
\begin{equation}
  R_0 \simeq 6.701 \,\mathrm{km} \,.
\end{equation}
The typical size scales to be expected from our Hartree-Fock theory must thus be of this order 
of magnitude, which corresponds well to the typical observed size of neutron stars 
of the order of $10 \, \rm{km}$. \\
With those parameters we have solved~\eqeqref{eq:HF1diff} for the condensate density in the inner regime using 
the boundary conditions~\eqref{eq:HF1bound} and subsequently employed~\eqeqref{eq:nth1} to calculate 
the thermal density in the inner regime. With the boundary conditions~\eqref{eq:bc1} and~\eqref{eq:bc2} at 
$R_0$ and quantities like $N_{\mb{0}}$ and $N_{\mathrm{th,1}}$ extracted from the inner solution, we then 
continue to solve~\eqeqref{eq:nthoutFinal} for $H(r)$ and obtain the thermal density in the outer regime 
from~\eqeqref{eq:hnth}. In~\figref{fig:densityprofiles} we show the corresponding solutions for both of the 
densities for a range of temperatures from $10^{11} \,\mathrm{K}$ to $4\cdot 10^{11}\, \mathrm{K}$ and for the 
total number of particles $N_{\mathrm{tot}}=10^{57}$. The condensate is given by the black curve, whereas the 
thermal density is plotted in red. For all simulations, the central density $\rho_c$ and the Thomas-Fermi 
radius $R_0$ are shown in~\tabref{tab:summary}. We have also listed the corresponding thermal radius $R_{\mathrm{th}}$ 
which denotes the border of the star, i.e. the point where the thermal density in the outer regime has 
fallen off to zero.

\section{Astrophysical implications} 
\label{sec:astro}
We will now proceed to extract results from the above calculations which are of astrophysical 
relevance, deducing various macroscopic and observable quantities for neutron stars. 

The parameters we will consider are the mass and the radius of the neutron star. Furthermore, 
we determine a restriction on the possible masses in form of a maximum allowed mass derived from physical 
constraints, and the equation of state of the neutron star. \\
In general, the determination of neutron star properties from observations is not straightforward. 
It differs from case to case and often involves the deduction of parameters from a combination of 
directly observable parameters or even assumptions on the physics inside the star. Neutron stars are 
rotating, magnetized objects, which can exist on their own or as part of a binary system. Its magnetic 
fields usually lead to the emission of electromagnetic radiation at the magnetic poles of the star. 
If the emitted beam lies in the direction of the earth, it is possible to detect this radiation, which 
pulsates with the frequency of the star's rotation, and, if the neutron star is part of a binary system, 
is further modulated with the orbital period of the binary. Due to this pulsed emission, neutron stars 
are also called pulsars. The emission can lie in a broad range of frequencies, from radio via optical 
to X-ray and $\gamma$-ray frequencies, depending on the specific properties of the star itself and on 
the possible companion star. \\
The physical observables of neutron stars are few. Besides the spectra detected from the neutron star 
and its companion, observations of the rotation, and the orbit in a binary system, are the most important 
features. \\
In the case of isolated neutron stars the spectrum can be very insightful since it is not contaminated 
by the influence of a companion or the remnants of a supernova. From the spectroscopy of the detected 
radiation and the timing of the pulses and their redshift, it is possible to infer temperature and distance 
to the observer, 
which yields the star's radius. From certain emission features in the spectrum, it might also be possible 
to deduce the gravitational redshift at the surface of the star, which constrains the relation of the mass 
to the radius~\cite{1982Lamb} -- and thus even the mass of the neutron star can be obtained. In a binary 
system on the other hand, where the neutron star accretes material from its companion, the X-ray bursts 
from the accretion process can be fit to a black body spectrum and thus, via temperature, flux and distance 
of the binary system, the radius of the star is obtained as well~\cite{1987Para}. \\
Besides the spectrum, the orbital parameters of a neutron star in a binary system are crucial in order 
to estimate its mass. Some neutron stars feature planetary systems, which lead to the determination of the 
neutron star's mass via Kepler's laws of planetary motion~\cite{1992Wols}. About 5\% of neutron stars are 
part of a binary system - in these cases, the exact observation of the companion can yield important 
information on the neutron star's properties. Via the Keplerian laws and the law of gravitation the masses 
of the neutron star and its companion can be expressed in terms of parameters like the orbital 
period, the radial velocities and the inclination angle of the orbit with respect to the line of sight to 
the observer~\cite{2007Haen}. The radial velocities can in turn be obtain from the measurement of the 
Doppler shifts of the spectra. Depending on how many parameters can be successfully determined from 
observations of the orbit, and how much additional information can be extracted from the spectra, 
one or both of the masses of the binary system can be calculated. The mass of the most massive 
neutron star found so far was calculated from orbital parameters and the mass of the white dwarf companion, 
obtained from the spectroscopy of the detected energy spectrum~\cite{2013Anto}. In some cases, in particular 
for radio pulsar binaries with very compact orbits, the orbital parameters can be determined with such 
precision that the detection of general relativistic effects is possible~\cite{2007Latt}. 
The mass of another very massive pulsar was thus determined using Shapiro delay, a gravitational time delay 
effect on the radiation of the pulsar due to the presence of the companion~\cite{2010Demo}. \\
In cases where a clean calculation of the star's radius from the spectrum is not possible, the radius is 
often inferred from the determination of the mass and assumptions on the star's density, which is believed 
to be of the order of nuclear density. Due to the unknown nature of neutron star's interiors, and the fact 
that the equation of state of neutron stars can unfortunately not be measured directly, these radius estimates 
are however highly uncertain. \\
The equation of state of neutron stars is subject to wide speculation and has spawned many different models 
describing the physical processes inside a neutron star. Many models assume a composition of nuclear or 
neutron matter, but this assumption still admits a broad range of possible equation of states. Also more exotic 
models with other particle species have been discussed, as already mentioned in the introduction. The impact of 
the equation of state is mirrored e.g. in the mass-radius relation, and can be constrained from observations 
if both mass and radius are reliably known. Also the distribution of neutron star masses from an ensemble of 
observations can give clues on the equation of state, by comparing the maximally allowed masses predicted by 
a certain equation of state with the maximum masses of neutron stars found in observations. We will employ the 
latter method to compare our calculations to observational information.

\subsection{Mass and density plots}
The total mass of the star in our model is given by 
\begin{equation}
  M = 4\pi m \int_0^{R_{\mathrm{th}}} dr\, r^2\, \Big[ n_{\mb{0}}(r) + n_{\mathrm{th}}(r) \Big] \,,
\end{equation}
obtained via the numerical integration of the respective density profiles and multiplication with the mass 
$m$ of a neutron pair. Our simulations were carried out for the example of $N_{\mathrm{tot}} = 10^{57}$, which 
corresponds to a mass of $M \simeq 1.7 \, M_{\odot}$. We can obtain density profiles and thus objects with 
arbitrarily high mass by modifying $A$, which determines the total number of particles. It is not possible 
to obtain an upper limit on the mass from our calculations since the simulations can be carried out for an 
arbitrary number of particles. Therefore we have to resort to other methods to obtain a limitation of the 
mass, employing either the general relativistic limit, i.e. the Schwarzschild limit of gravitational 
collapse, or an upper bound on the speed of sound of the particles inside the star, demanding that 
causality may not be violated. It turns out that in the case of a neutron star the Schwarzschild 
limit yields a more stringent condition than the limit on the speed of sound. The Schwarzschild limit requires 
the object to be larger than its Schwarzschild radius to prevent gravitational collapse into a black hole, i.e., 
\begin{equation}
  R_{\mathrm{th}} > r_S = \frac{2 G M}{c^2} \,.
\end{equation}
For the simulation with $N_{\mathrm{tot}} = 10^{57}$ particles, i.e. a mass of $M \simeq 1.7 M_{\odot}$, the 
Schwarzschild radius turns out to be $r_S \simeq 3.17\, \mathrm{km}$, which is below the obtained thermal radii of the 
configurations, see~\tabref{tab:summary}. However, it is possible to turn around the criterion and calculate 
the maximum possible mass for the size scales obtained in our simulations, via 
\begin{equation} ~\label{eq:Mmax0}
  M_{\mathrm{max}} = \frac{c^2 R_{\mathrm{th}}}{2 G} \,.
\end{equation}
By using the dependence of the thermal radius on the temperature as obtained from the numerical results, 
it is possible to obtain a limit on the maximum mass of the system as a function of temperature. We will 
elaborate further on this issue in~\secref{sec:Mmax}. 

\subsection{Size scales}
\begin{figure} 
	\centering
	\includegraphics[width=0.5\textwidth]{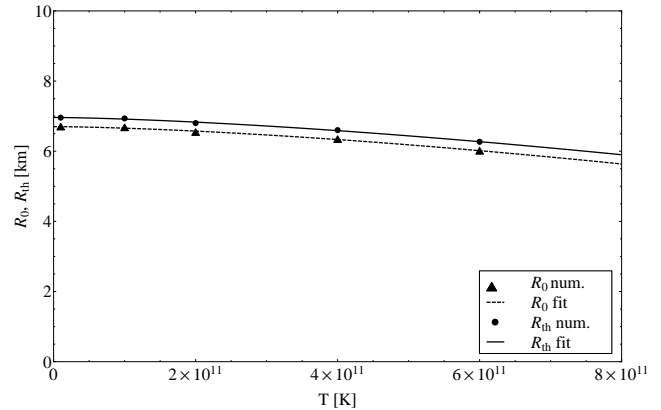}
	\caption{Dependence of the Thomas-Fermi radius $R_0$ and the thermal radius $R_{\mathrm{th}}$ 
	on temperature $T$: results from the simulations (dots, triangles) and numerical fits (solid, dashed) 
	as given by~\eqsref{eq:radiusfit0}--\eqref{eq:Rthfit}. }
	\label{fig:R}
\end{figure}

Besides the mass, another quantity of interest is the size of the system. We represent the condensate 
radius $R_0$ and the total radius $R_{\mathrm{th}}$ of the star in~\figref{fig:R} as a function of 
temperature. The dots and triangles give the numerical results obtained in the simulations, and 
the curves show the best fit for the numerical data. For the condensate radius, the general form 
\begin{equation} \label{eq:radiusfit0}
    R_0 (\theta) = R_0 + a_1 \, \theta^{a_2} \,
\end{equation}
was used for the fit, where $R_0 = \pi/\sigma = 6.701 \,\mathrm{km}$  is the Thomas-Fermi radius for zero 
temperatures, and $\theta$ is the dimensionless temperature. The best fit results yield 
\begin{equation} \label{eq:R0fit}
  a_1 = -0.342 \,\mathrm{km} \,,\quad a_2 = 1.53 \,.
\end{equation} 
The fitting ansatz for the thermal radius was 
\begin{equation} \label{eq:radiusfitTh}
    R_{\mathrm{th}} (\theta) = b_1 + b_2 \, \theta^{b_3} \,
\end{equation}
where the results read 
\begin{equation} \label{eq:Rthfit}
  b_1 = 6.962 \, \mathrm{km}\,,\quad~ b_2 = - 0.349 \,\mathrm{km} \,, \quad b_3 = 1.5001 \,. 
\end{equation}
Both exponents, in particular the one for the thermal radius, are very close to the value $1.5$, which 
can be ascribed to the leading dependence of any occurring variable on the temperature to the power of 
$3/2$. Any deviations from the exact power $3/2$ stem from the argument of the polylogarithmic function, 
which contains a further dependence on the temperature. 
We see that not only the condensate radius is decreasing with rising temperatures, but also the thermal 
radius, despite the growing expansion of the thermal cloud at the border of the star. In total the star 
is thus decreasing in size with rising temperatures, while its central density increases correspondingly. 
The size scales are of the order of $6\, \mathrm{km}$, which is determined by the zero-temperature limit and 
only depends on the natural constants $G$ and $\hbar$ and the choice of the parameters $m$ and $a$. 

\subsection{Maximum mass}
\label{sec:Mmax}
\begin{figure}
	\centering
	\includegraphics[width=0.5\textwidth]{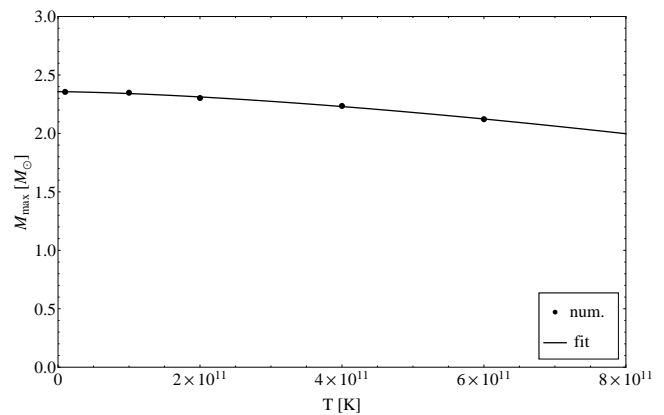}
	\caption{Maximum mass as a function of temperature, as inferred from the limits given 
	by the Schwarzschild criterion: numerical results (dots) and a fit (solid) as given 
	by~\eqsref{eq:EoSMmaxFit} and~\eqref{eq:EoSMmaxFitValues}. }
	\label{fig:MmaxT}
\end{figure}

\begin{figure}
	\centering
	\includegraphics[width=0.5\textwidth]{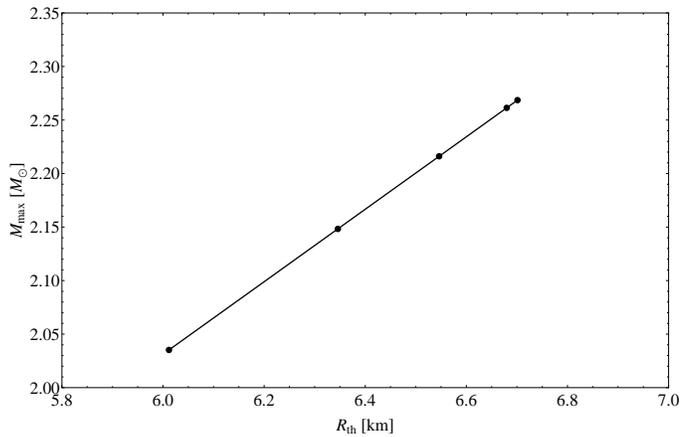}
	\caption{Maximum mass plotted over the thermal radius $R_{\rm{th}}$, for different 
	values of the temperature $T$.}
	\label{fig:MR}
\end{figure}

Subsequently, we can proceed to derive a maximum mass for the system by employing the upper limit on the 
mass as given by the Schwarzschild limit. Generalizing~\eqeqref{eq:Mmax0} to finite temperatures, we 
obtain 
\begin{equation} \label{eq:MmaxT}
  M_{\mathrm{max}} = \frac{2R_{\mathrm{th}}(T)}{G c^2} \,,
\end{equation}
and employing the temperature dependence of the thermal radius as given by~\eqsref{eq:radiusfitTh} 
and~\eqref{eq:Rthfit}, we can compute the correspinding mass limits for the system, shown in~\figref{fig:MmaxT}. 
The limit for zero temperatures can be computed employing the Thomas-Fermi radius $R_0$ as given 
by~\eqeqref{eq:TFT0analytic}, 
\begin{equation} \label{eq:MmaxT0}
  M_{\mathrm{max},0} = \frac{2 R_0}{G c^2} = \frac{\pi \hbar c^2 \sqrt{a}}{2 (G m)^{3/2}} \,,
\end{equation}
and results in the value $M_{\mathrm{max},0} \simeq 2.3  M_{\odot}$. 
The qualitative temperature dependence of $M_{\mathrm{max}}$ can be inferred again from a fit of the 
curve with a general fitting function 
\begin{equation} \label{eq:EoSMmaxFit}
  M_{\mathrm{max}}(\theta) = M_{\mathrm{max},0} + d_1 \, \theta^{d_2} \,,
\end{equation}
resulting in the best fit values 
\begin{equation} \label{eq:EoSMmaxFitValues}
  d_1 = - 0.118 \,M_{\odot} \,, \quad d_2 = 1.5001\,.
\end{equation}
Again, we obtain a small, but distinct dependence on the temperature to the power of $3/2$. The 
maximum mass $2.3 \, M_{\odot}$ for zero temperatures is larger than the original limit on 
neutron stars given by Tolman, Oppenheimer and Volkoff \cite{1939Tolm,1939Oppe} and corresponds 
well to observational evidence \cite{2010Demo,2013Kizi}. The decrease of the maximum possible mass 
with increasing temperatures can be understood by considering the increase in the central condensate 
density with higher temperatures - the condensate seems to be compressed by the thermal density, 
which makes the object smaller and thus leads to a smaller mass given by the Schwarzschild limit. 
This is supported by the results for the equation of state of the condensate, as computed in the next 
subsection. \\
For neutron stars, a commonly shown plot is the relation between maximum mass and radius. In our 
model, we obtain a mass-radius-relation by plotting the $(M,R_{\rm{th}})$-pairs for the different 
temperatures used in the computations, shown in~\figref{fig:MR}. As expected from~\eqeqref{eq:MmaxT}, 
the dependence of $M_{\mathrm{max}}$ on $R_{\rm{th}}$ is linear, and thus the plot shows no peculiar 
structure. This is due to the imposition of the Schwarzschild criterion to calculate the maximum 
allowed masses, instead of having a natural maximum mass limit given by an instability of 
the theoretical description.

\subsection{Equation of state} 
\label{sec:EoScS}
\begin{figure}
	\begin{center}
	\subfigure[~Condensate equation of state. \label{fig:EoScond}]{\includegraphics[width=0.5\textwidth]{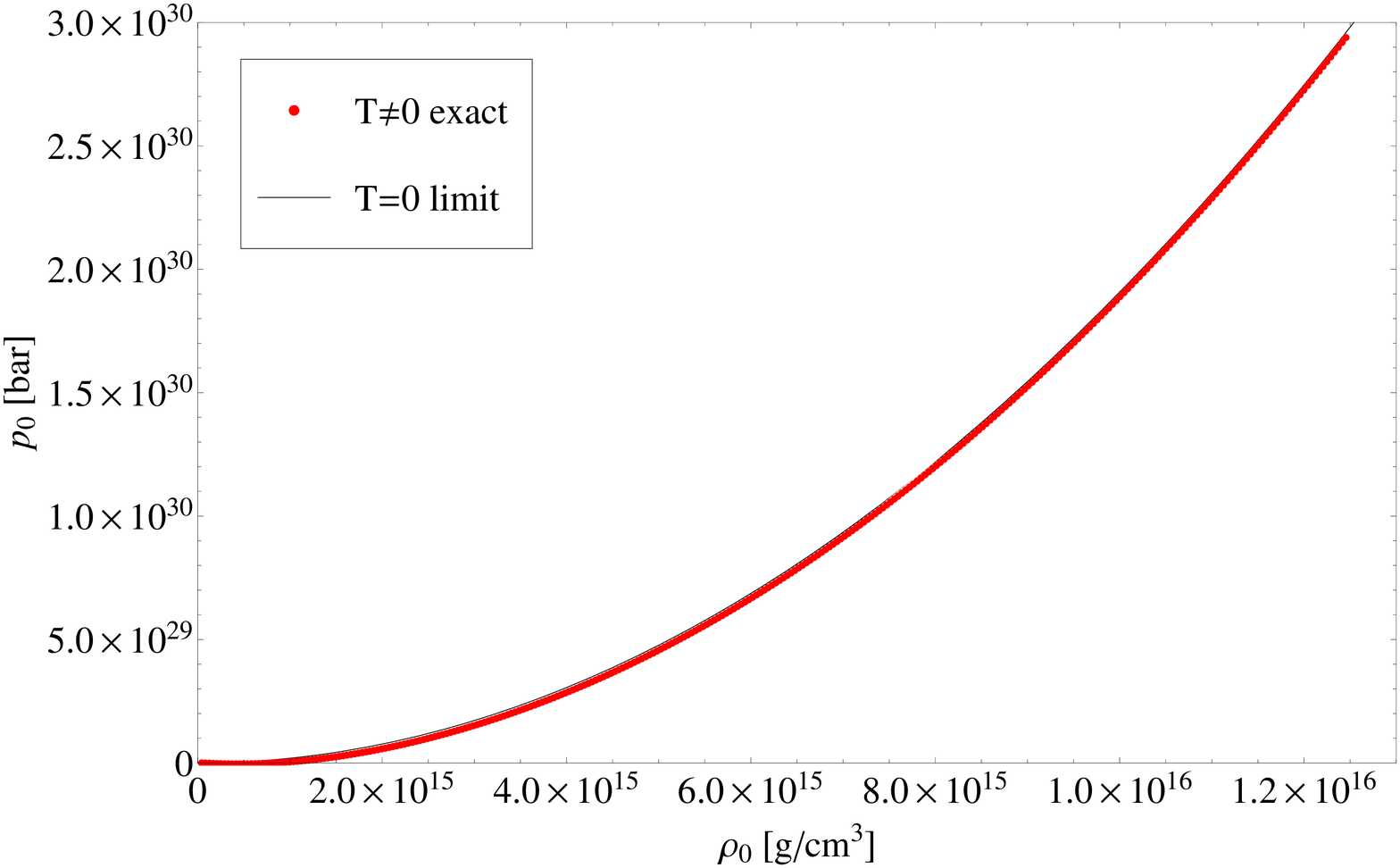}}
	\subfigure[~Condensate equation of state in a close-up. \label{fig:EoScondCloseUp}]{\includegraphics[width=0.5\textwidth]{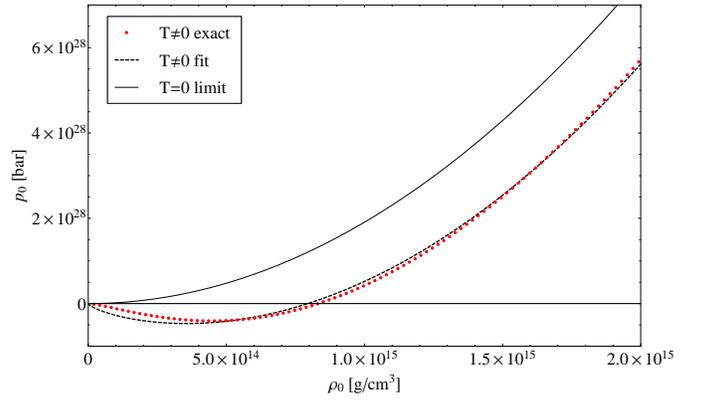}}
	\caption{Equation of state of the condensate for the example of $T = 4\cdot 10^{11} \mathrm{K}$ and 
	$N_{\mathrm{tot}} = 10^{57}$, as obtained from the exact formulation~\eqref{eq:pcond} (dots) and the 
	equation of state for the zero-temperature limit~\eqref{eq:EoSwithrho} (dashed). Both curves are 
	rather close, so that they cannot be distinguished in~\figref{fig:EoScond}, whereas in~\figref{fig:EoScondCloseUp} 
	a close-up for small densities is shown, where the discrepancy is noticable. A numerical fit (solid) as given 
	by~\eqsref{eq:EoSCondFit} and~\eqref{eq:EoSCondFitValues} was performed. At the density 
	$\rho \simeq 7.95 \cdot 10^{14} \mathrm{g}/\mathrm{cm}^3$ the pressure becomes negative. }
	\label{fig:EoS}
	\end{center}
\end{figure}

Ultimately, we investigate the equation of state, i.e. the characteristic relation of 
pressure and density $p=p(\rho)$ of the matter in the star. In principle, a BEC has 
a polytropic equation of state with an index $n=1$, which is an equation of state that has been 
used in the context of neutron stars before~\cite{2011Cha2}. However, since in our 
system two different phases of matter coexist, we have to define an equation of state for each of them 
independently. In the case of thermal fluctuations, we further have to consider the two different regimes 
inside and outside of the Thomas-Fermi radius. Thus we have to distinguish three phases of matter with 
different equations of state. \\
The equation of state of the condensate was derived in Refs. \cite{2011Cha2,2012Hark} for a system obeying the 
same Hamiltonian as given in~\eqeqref{eq:Heisenberg}. Adding a small perturbation to the mean field wave function 
of the condensate and using a Madelung ansatz for the mean field itself, it is again possible to derive a set of 
hydrodynamic equations, i.e. the continuity and Euler equations, from the Heisenberg equation, but this time under 
the inclusion of thermal fluctuations. From Eq.~(40) in Ref. \cite{2012Hark} the gradient of the pressure can be 
read off by comparison to a general Euler equation for a hydrodynamic system as 
\begin{equation} \label{eq:nablaP}
  \nabla p_{\mb{0}} = n_{\mb{0}} \, \nabla \left[ g \left( n_{\mb{0}} + 
	2 n_{\mathrm{th}} \right) \right] \,.
\end{equation}
Subsequently we can calculate the pressure of the condensate by integrating~\eqeqref{eq:nablaP}. 
This leads to the well-known polytropic equation of state for the pure condensate with polytropic index $n=1$, 
and a correction term proportional to a polylogarithm of order $5/2$, as well as a term containing both 
condensate and thermal fluctuations, and a constant, 
\begin{eqnarray} \label{eq:pcond}
  p_{\mb{0}} &=& \frac{g}{2m^2} \,\rho^2 + \frac{2}{\beta \lambda^3} \, 
    \zeta_{5/2} \big[ e^{- \frac{\beta g}{m} \rho} \big] \\
  &~&~~~~~ + \frac{2g}{m^2\, \lambda^3} \,\rho \,
    \zeta_{3/2} \big[ e^{- \frac{\beta g}{m} \rho} \big] - \frac{2}{\beta \lambda^3} \,\zeta_{5/2}(1) \,. \nonumber
\end{eqnarray}
Here again $\rho = m\, n_{\mb{0}}$ denotes the mass density of the condensate. Equation~\eqref{eq:pcond} is the 
equation of state for the condensate with corrections from the thermal density. \figref{fig:EoScond} shows the 
condensate pressure given as a function of the condensate density for the example of $T = 4\cdot 10^{11} \mathrm{K}$ 
and $N_{\mathrm{tot}} = 10^{57}$. As we can see from the close-up of the condensate equation of state 
in~\figref{fig:EoScondCloseUp}, the pressure turns out to become negative for small densities. 
This is a consequence of the Thomas-Fermi approximation for the condensate: at the border of the star, where 
the condensate density is small, the quantum pressure of the condensate, which we had neglected, becomes important. 
For the small densities at the border of the star, the quantum pressure would thus correct the unphysical negative 
pressures obtained in~\eqref{eq:pcond}. Considering this correction, the pressure of the condensate would presumably 
increase for small densities, which would explain the compression of the condensate and subsequent shrinking of the 
star with increasing temperatures, as obtained in the previous subsections. \\
Besides the exact form of the condensate pressure~\eqref{eq:pcond}, denoted by the dots, and the 
zero-temperature limit (dashed), \figref{fig:EoScondCloseUp} contains a fit (solid), carried out with the 
general polytropic ansatz for the pressure as a function of the dimensionless condensate density $\tilde{n}_{\mb{0}}$, 
\begin{equation} \label{eq:EoSCondFit}
  p_{\mb{0}} =  p_{\mb{0}}^{(0)}\,\tilde{n}_{\mb{0}}^2 + c_1 \,\tilde{n}_{\mb{0}}^{c_2} \,.
\end{equation}
where the coefficient for the first term is 
\begin{equation}
  p_{\mb{0}}^{(0)} = \frac{1}{2}\frac{(k_B T_{\mathrm{ch}})^2}{g} \,,
\end{equation}
and for a neutron star with the chosen specifications amounts to $p_{\mb{0}}^{(0)} = 3.288\cdot 10^{27}\, \mathrm{bar}$. 
The best fit for the parameters $c_1$ and $c_2$ resulted in the values 
\begin{equation} \label{eq:EoSCondFitValues}
  c_1 = -0.105\,\mathrm{bar}  \,, \quad\quad\quad~ c_2 = 0.703 \,.
\end{equation}
The parameter $c_2$ in the exponent leads to the polytropic index 
\begin{equation}
  n_2 = - 3.363 \,,
\end{equation}
which implies that the polytropic form with $n=1$ for the condensate at $T=0$ is modified at finite temperatures 
to obtain another polytropic component with negative index $n_2$, which is due to the presence of the thermal 
density. Negative polytropic indices denote metastable states of matter which can occur in highly energetic 
processes and environments in astrophysics \cite{2004Hore}. Since the thermal cloud makes up only a small fraction 
of the total number of particles however, as can be seen from the respective smallness of $c_1$ as compared to 
$p_{\mb{0}}^{(0)}$, and moreover negative pressures only occur for very small densities of the order of less than 
$10^{15}\, \mathrm{g}/\mathrm{cm}^3$ at the border of the star, we infer that the negative polytrope component does not 
endanger the stability of the system as a whole. We have calculated the percentage of the Thomas-Fermi radius for 
which the pressure becomes negative, which happens at the density $\rho \simeq 7.95 \cdot 10^{14} \mathrm{g}/\mathrm{cm}^3$. 
For the example of $T = 4\cdot 10^{11} \,\mathrm{K}$ this corresponds to the radius $r = 6.346 \, \mathrm{km}$, which is 
equivalent to $0.99997\,R_0$. \\
%
For the thermal cloud, the pressure can be obtained from its definition 
\begin{equation}
	p_{\mathrm{th}}(r) = \int \frac{d^3k}{(2\pi)^3} \frac{\hbar^2 \mb{k}^2 / 2m}{e^{\beta 
		\left[ \epsilon_{\mb{k}}(r) - \mu \right]} -1} \,,
\end{equation}
which leads to a polylogarithmic function, similar to the thermal density, but with an index $5/2$: 
\begin{equation}
	p_{\mathrm{th}}(r) = \frac{1}{\beta \lambda^3} \,\zeta_{5/2} \left[ e^{-\beta \left( 2g \,
		\left[n_{\mb{0}}(r)+n_{\mathrm{th}}(r)\right] + \Phi(r) - \mu \right) } \right] \,.
\end{equation}
For the two regimes, we can formulate the pressure as 
\begin{eqnarray} \label{eq:pth}
	p_{\mathrm{th},1}(r) &=& \frac{1}{\beta \lambda^3} \,\zeta_{5/2} \left[ 
		e^{ -\beta g n_{\mb{0}}(r)} \right] \,, \label{eq:pth1} \\
	p_{\mathrm{th},2}(r) &=& \frac{1}{\beta \lambda^3} \,\zeta_{5/2} \left[ e^{-\beta \left( 2g \,
		n_{\mathrm{th},2}(r) + \Phi(r) - \mu \right) } \right] \,.\label{eq:pth2}
\end{eqnarray}
The results can be obtained in analogy to the solution for the thermal density in the respective regimes. 
For thermal fluctuations, the functional dependence as given by results~\eqref{eq:pth1} and~\eqref{eq:pth2} 
is exactly what is to be expected for a thermal gas of bosons, and confirms the vanishing pressure of free 
bosons for zero temperatures. \\
However, when attempting to numerically compute the thermal pressure from~\eqref{eq:pth1} and~\eqref{eq:pth2} 
we run into problems, as the pressure becomes complex around the border of the condensate. This deficiency 
is again due to the Thomas-Fermi approximation, which affects the condensate and thus also the thermal 
fluctuations in that regime, and the polylogarithmic function, which becomes complex for arguments larger 
than one. Therefore we could not obtain numerical results for the thermal pressure, which only 
differs significantly from zero in the critical region at the border of the condensate.

\section{Conclusions and Outlook}
\label{sec:Concl}
The work presented in this paper investigated the occurrence of a BEC phase in compact astrophysical objects 
such as neutron stars. A careful consideration of the typical environments showed that the neutrons inside 
neutron stars are likely to form pairs due to the strong nuclear forces between them, similarly to an atomic 
nucleus, i.e. are present in a superfluid state. These neutron pairs are considered as the effective 
bosonic elementary particle in the BEC.
The model presented in this article starts from this simplified picture of very strongly bound neutron 
pairs as perfect bosons, and does not take into account the presence of single neutrons or 
other particle species. 
Our work represents a first step towards an alternative description of neutron stars based on the phenomenon 
of BCS-BEC-crossover in nuclear or neutron matter, and increasing efforts by theoreticians to consider these 
scenarios validate our efforts to compute observable quantities that can be compared to observations. \\
We would like to emphasize though that a physically more exact treatment would require the investigation 
of the BCS-BEC crossover itself along the lines of Refs. \cite{2005Astr,2006Mats,2007Marg,2013Sala}, not just 
the BEC limit. The full crossover would unify the different physical behaviour of the BCS and BEC regimes into 
one theory, and would apply to both fermion and boson stars simultaneously in the respective limits of the 
theory. The treatment we have set up must result from the complete crossover theory as the BEC limit, and should 
thus only be regarded as an approximate solution to the issue. \\
In the BEC limit, the system was treated within the framework of a Hartree-Fock theory, starting from a 
Hamiltonian including contact and gravitational interactions between the particles. Self-consistency 
equations determining the wave functions of condensate and thermal fluctuations were obtained from the 
variation of the free energy of the system. In analogy to these derivations, the semi-classical limit 
of both the free energy and the Hartree-Fock equations was formulated, describing the system in terms of 
the densities of condensate and thermal fluctuations. The resulting equations were processed further up to 
a certain point, before the solutions for both the profiles of condensate and thermal density as a function 
of the radial distance from the center of the star were obtained by numerical procedures. Integrating out 
the obtained densities leads to the total mass of the system, along with other quantities of astrophysical 
consequences. 
From our model, we have obtained objects with radii of about $6\, \mathrm{km}$, masses of about $2.3\, M_{\odot}$ 
and central densities around $\rho_c \simeq 10^{16} \mathrm{g}/ \mathrm{cm}^3$, which approximately coincide with 
the typical values to be expected for neutron stars. Since from the zero-temperature limit and the subsequent 
analysis for finite temperatures, the radial extension of the system was found to be around $6\, \mathrm{km}$, 
decreasing with a temperature dependence proportional to $T^{3/2}$, we were able to employ the 
Schwarzschild criterion of gravitational collapse in order to derive a mass limit on the neutron stars, 
which lead to a maximum mass of about $2.3\, M_{\odot}$, decreasing proportional to $T^{3/2}$ as well. The 
order of magnitude of these results seems plausible considering observational evidence. \\
As already stated at the outset, the theory contains several simplifications, introduced in order to make 
the system more treatable. Some of them were mathematically motivated, whereas others have been general 
physical assumptions within our model from the beginning. We considered a phenomenon mainly known from 
ultracold quantum gases in laboratory scenarios and applied an established mathematical treatment to a 
rather unusual field of application, namely the large scales of astrophysics. It is therefore to be expected 
that simplifications and idealizations are necessary in order to obtain results. \\
On the mathematical side, we have carried out a Hartree-approximation for the gravitational part of the 
interactions, which eliminated the bilocal Fock terms in the expressions. The inclusion of these 
terms could perhaps be treated in form of an appropriate local density approximation. \\
The theory is limited to low temperatures, where by definition the particles in the thermal phase 
are few and the condensate dominates. However, the necessity to develop a more complete theory featuring a 
smoother description of the high-temperature transition region between condensate and thermal state of the 
system, incorporating the breakdown of the condensate as a phase transition, is obvious. \\
A further assumption of the theory is a spatially constant temperature throughout the star, which is unlikely 
to hold in realistic physical situations. This is closely connected to the breakdown of the condensate 
towards the outer layers of the star, where the density and thus the critical temperature decrease, and at a 
certain point the condition $T < T_{\mathrm{crit}}$ for the formation of a condensate cannot be met anymore. The 
inclusion of spatial variation of temperature in the self-consistency equations would thus allow for a much more 
detailed and realistic model. \\
Finally, we would like to comment on the possibility of rotation. It is presumed that most of the compact objects 
in the universe rotate, since an evolution of a completely static system is highly unlikely in an initially hot 
and violent universe. Rotation of BECs in laboratory environments have been shown to exhibit new phenomena like 
the formation of vortices of normal phase matter inside the BEC \cite{2009Fett}, growing with increasing temperature 
until the breakdown of condensate at the transition to the thermal phase. The existence of a vortex in a Bose star, 
or, more realistically, a grid of vortices, should be assumed, which grow in width and finally cause a transition 
to a normal phase Bose star with increasing temperature. The inclusion of rotation is expected to lead to a 
destabilization of the system due to the presence of tidal forces, and thus should lead to a higher maximum mass 
counterbalancing the increased outwards forces. Further, rotation could potentially help to explain 
dynamical phenomena observed in neutron stars, like e.g. glitches in the rotation frequency, and would provide further 
means to compare our results to observations. \\
Thus there is a large number of possibilities to generalize and extend the present work.

\section*{Acknowledgments}
We would like to thank James Anglin, Hagen Kleinert, Jutta Kunz, Jorge Rueda and Remo Ruffini for useful discussions, 
as well as the Collaborative Research Center SFB/TR 49 of the German Research Foundation (DFG) for the support during 
the last stages of the work. Part of the work of C.G. was supported by the Erasmus Mundus Joint Doctorate Program by 
Grant Number 2010-1816 from the EACEA of the European Commission.

\bibliographystyle{unsrt}
\bibliography{bibliography.bib}

\appendix
\section{Hartree-Fock theory for bosons}
\label{sec:HF}
In the appendix, we derive the Hartree-Fock theory 
at finite temperatures for a generic system of bosons, employing the formalism of the grand-canonical 
ensemble and its definition of the free energy. By means of a variational principle we then determine 
a set of coupled self-consistency equations for the wave functions of both condensate and thermal 
fluctuations. The derivation relies largely on the formalism introduced in Ref. \cite{1997Oehb}, and has 
been adapted for our scenario.

\subsection{Free energy}
We start from the general Hamiltonian 
\begin{eqnarray} \label{eq:HamGen}
  && \hat{\mathcal{H}} = \int d^3x \,\hat{\Psi}^{\dag}(\mb{x}) \bigg[ h(\mb{x}) - \mu \\
  && ~~~~~ + \frac{1}{2} \int d^3x' \, \hat{\Psi}^{\dag}(\mb{x}') U(\mb{x},\mb{x}') 
    \hat{\Psi}(\mb{x}') \bigg] \hat{\Psi}(\mb{x}) \,, \nonumber
\end{eqnarray}
where the first-quantized Hamiltonian operator $h(\mb{x})$ is defined as the kinetic term 
plus an external potential, 
\begin{equation}
  h(\mb{x}) = -\frac{\hbar^2}{2m} \Delta + V(\mb{x})\,,
\end{equation}
and the interaction term $U(\mb{x},\mb{x}')$ is as yet unspecified. The field operators $\hat{\Psi}^{\dag}$ 
and $\hat{\Psi}$ obey the usual commutator relations for bosonic particles, 
\begin{eqnarray}
  && \left[ \hat{\Psi}^{\dag}(\mb{x}), \hat{\Psi}^{\dag}(\mb{x}') \right] = 
   \left[ \hat{\Psi}(\mb{x}), \hat{\Psi}(\mb{x}') \right] = 0 \,, \\
  && \left[ \hat{\Psi}(\mb{x}), \hat{\Psi}^{\dag}(\mb{x}') \right] = \delta(\mb{x}-\mb{x}') \,. 
\end{eqnarray}
The grand-canonical formalism defines the partition function $Z$ as 
\begin{equation} \label{eq:partfunc}
  Z = \Tr \left[ e^{-\beta \hat{\mathcal{H}} } \right] \,,
\end{equation}
where $\beta=1/ (k_{\mathrm{B}} T)$ is the inverse temperature 
and the trace in the expression has to be taken over all states of the Fock space. \\
We now derive the equations that govern the state of the field operators. To this 
purpose, we employ a for now unknown one-particle basis $\Psi^{}_{\mb{n}}(\mb{x})$ characterized by  
discrete quantum numbers $\mb{n}$, and write the field operator as an expansion with respect 
to these functions $\Psi^{}_{\mb{n}}(\mb{x})$ as 
\begin{equation}
  \hat{\Psi}(\mb{x}) = \sum_{\mb{n}} \hat{a}^{}_{\mb{n}} \, \Psi^{}_{\mb{n}}(\mb{x}) \,, 
  \hat{\Psi}^{\dag}(\mb{x}) = \sum_{\mb{n}} \hat{a}^{\dag}_{\mb{n}} \, \Psi^{*}_{\mb{n}}(\mb{x}) \,.
\end{equation}
The expansion coefficients $\hat{a}^{\dag}_{\mb{n}}$ and $\hat{a}^{}_{\mb{n}}$ represent the 
creation and annihilation operators of a particle with the quantum number $\mb{n}$, and they 
obey similar commutator relations as the field operators $\hat{\Psi}^{\dag}$ and $\hat{\Psi}$ 
above. The one-particle basis is chosen to be orthonormal and thus 
\begin{eqnarray} \label{eq:normcomm}
  && \int d^3x \,\Psi^{*}_{\mb{n}}(\mb{x}) \, \Psi^{}_{\mb{n}'}(\mb{x}) = \delta_{\mb{n},\mb{n}'} \,,\\
  && \sum_{\mb{n}} \Psi^{*}_{\mb{n}}(\mb{x}) \, \Psi^{}_{\mb{n}}(\mb{x}') = \delta(\mb{x}-\mb{x}')
\end{eqnarray}
hold. We can then write the Hamiltonian operator~\eqref{eq:HamGen} in terms of these creation and 
annihilation operators as 
\begin{eqnarray}
  && \hat{\mathcal{H}} = \sum_{\mb{n}} \sum_{\mb{n}'} E^{}_{\mb{n},\mb{n}'}\, \hat{a}^{\dag}_{\mb{n}} 
    \hat{a}^{}_{\mb{n}'} \\
  && ~~~~~ + \frac{1}{2} \sum_{\mb{n}} \sum_{\mb{m}} \sum_{\mb{m}'} \sum_{\mb{n}'} 
    U^{}_{\mb{n},\mb{m},\mb{m}',\mb{n}'} \,\hat{a}^{\dag}_{\mb{n}} \hat{a}^{\dag}_{\mb{m}} 
    \hat{a}^{}_{\mb{m}'} \hat{a}^{}_{\mb{n}'} \,, \nonumber
\end{eqnarray}
where the respective matrix elements read 
\begin{eqnarray}
  && E^{}_{\mb{n},\mb{n}'} = \int d^3x \,\Psi^{*}_{\mb{n}}(\mb{x}) \left[ h(\mb{x}) - \mu \right] 
    \Psi^{}_{\mb{n}'}(\mb{x}) \,,\label{eq:Enn} \\
  && U^{}_{\mb{n},\mb{m},\mb{m}',\mb{n}'} = \int d^3x \int d^3x' \,\Psi^{*}_{\mb{n}}(\mb{x}) 
    \Psi^{*}_{\mb{m}}(\mb{x}') \label{eq:Unmnm} \\
  && ~~~~~~~~~~~~~~~~~~~~~~~~~ \times \, U(\mb{x},\mb{x}') \, \Psi^{}_{\mb{m}'}(\mb{x}') 
    \Psi^{}_{\mb{n}'}(\mb{x}) \,. \nonumber
\end{eqnarray} 
To treat the system further, we suppose the existence of an effective Hamiltonian 
$\hat{\mathcal{H}}_{\mathrm{eff}}$ describing the system as effectively non-interacting 
with one-particle energies $\epsilon_{\mb{n}}$, i.e. 
\begin{equation} \label{eq:Heff}
  \hat{\mathcal{H}}_{\mathrm{eff}} = \sum_{\mb{n}} \left( \epsilon_{\mb{n}} - \mu \right) 
    \hat{a}^{\dag}_{\mb{n}} \hat{a}^{}_{\mb{n}} \,.
\end{equation}
Thus, the system is formulated in terms of an unknown one-particle basis $\Psi^{}_{\mb{n}}(\mb{x})$ 
with unknown one-particle energies $\epsilon_{\mb{n}}$. These quantities have been artificially introduced, 
which means that in the end the result should not depend on them. Inspired by variational perturbation 
theory \cite{2001HKVSF,2009Klei}, we now express the real Hamiltonian in terms of the effective Hamiltonian 
and an additional parameter $\eta$ as 
\begin{equation} \label{eq:Heta}
  \hat{\mathcal{H}}(\eta) = \hat{\mathcal{H}}_{\mathrm{eff}} + \eta \, \left( \hat{\mathcal{H}}
    -\hat{\mathcal{H}}_{\mathrm{eff}}\right) \,.
\end{equation}
If $\hat{\mathcal{H}}_{\mathrm{eff}}$ is a good approximation for the real Hamiltonian 
$\hat{\mathcal{H}}$, then the second term is small, and the grand-canonical partition 
function can be expanded into a Taylor series with respect to the difference of the 
two Hamiltonians. In the end, we have to set $\eta=1$ in order to obtain a valid 
identity in~\eqeqref{eq:Heta}. \\
Using relation~\eqref{eq:Heta}, the partition function~\eqref{eq:partfunc} can be written as 
\begin{equation} 
  Z(\eta) = \Tr \left\{ e^{-\beta \left[ \hat{\mathcal{H}}_{\mathrm{eff}} + \eta \, \left( \hat{\mathcal{H}}
    -\hat{\mathcal{H}}_{\mathrm{eff}}\right) \right] } \right\} \,.
\end{equation}
Expanding this expression into a Taylor series with respect to the assumed smallness of $\hat{\mathcal{H}}
-\hat{\mathcal{H}}_{\mathrm{eff}}$ leads to 
\begin{eqnarray}
  Z(\eta) &=& \Tr \left[ e^{-\beta \hat{\mathcal{H}}_{\mathrm{eff}} } \right] + 
      \left(-\beta \eta\right)\, \Tr \left[ \left( \hat{\mathcal{H}} 
      -\hat{\mathcal{H}}_{\mathrm{eff}} \right) 
      e^{-\beta \hat{\mathcal{H}}_{\mathrm{eff}} } \right] \nonumber\\
  && + \frac{1}{2} \left(-\beta \eta\right)^2 \, \Tr \left[ \left( 
      \hat{\mathcal{H}} -\hat{\mathcal{H}}_{\mathrm{eff}} 
      \right)^2 e^{-\beta \hat{\mathcal{H}}_{\mathrm{eff}} } \right] + ... \,. 
\end{eqnarray}
After defining the notions of the effective partition function 
\begin{equation} \label{eq:Zeff}
  Z_{\mathrm{eff}} = \Tr \left[ e^{-\beta \hat{\mathcal{H}}_{\mathrm{eff}} } \right] \,
\end{equation}
and the effective expectation value of an operator $\hat{X}$ as 
\begin{equation} \label{eq:effexpval}
  \langle \hat{X} \rangle^{}_{\mathrm{eff}} = \frac{1}{Z_{\mathrm{eff}}} \Tr \left[ \hat{X}\, e^{-\beta 
    \hat{\mathcal{H}}_{\mathrm{eff}} } \right] \,,
\end{equation}
we can rewrite the expansion of the partition function as 
\begin{eqnarray}
  Z(\eta) &=& Z_{\mathrm{eff}} \, \bigg[ 1+ \left(-\beta \eta\right) \langle \left( \hat{\mathcal{H}} 
    -\hat{\mathcal{H}}_{\mathrm{eff}} \right) \rangle^{}_{\mathrm{eff}} \\
  && ~~~~~ + \frac{1}{2} \left(-\beta \eta\right)^2 \langle \left( \hat{\mathcal{H}} 
    -\hat{\mathcal{H}}_{\mathrm{eff}} \right)^2 \rangle^{}_{\mathrm{eff}} + ... \bigg] \,. \nonumber
\end{eqnarray}
This is an expansion in terms of the moments, i.e. for the $n$\textsuperscript{th} order in the expansion 
the $n$\textsuperscript{th} power of the effective expectation value of $\left(\hat{\mathcal{H}}
-\hat{\mathcal{H}}_{\mathrm{eff}}\right)$ appears. The free energy 
\begin{equation}
  F(\eta) = -\frac{1}{\beta} \ln Z(\eta)
\end{equation}
can then be written as 
\begin{eqnarray} \label{eq:Fnu}
   F(\eta) &=& F_{\mathrm{eff}} - \frac{1}{\beta} \ln \bigg\{ 1 -\beta \eta \,\langle \left( \hat{\mathcal{H}} 
    -\hat{\mathcal{H}}_{\mathrm{eff}} \right) \rangle^{}_{\mathrm{eff}} \\
   && ~~~~~~~~~~ + \frac{1}{2} \,\beta^2 \eta^2 \langle \left( \hat{\mathcal{H}} 
    -\hat{\mathcal{H}}_{\mathrm{eff}} \right)^2 \rangle^{}_{\mathrm{eff}} + ... \bigg\} \,, \nonumber
\end{eqnarray}
with the effective free energy defined as 
\begin{equation} \label{eq:Feff}
  F_{\mathrm{eff}} = -\frac{1}{\beta} \ln Z_{\mathrm{eff}} \,.
\end{equation}
We then employ the Taylor expansion of the logarithm to expand the free energy~\eqref{eq:Fnu} into a series as 
\begin{eqnarray}
  && F(\eta) = F_{\mathrm{eff}} + \eta \,\langle \left( \hat{\mathcal{H}} 
    -\hat{\mathcal{H}}_{\mathrm{eff}} \right) \rangle^{}_{\mathrm{eff}} \\
   && ~~~~~ -\frac{1}{2} \, \beta \eta^2 \left[ \langle \left( \hat{\mathcal{H}} 
    -\hat{\mathcal{H}}_{\mathrm{eff}} \right)^2 \rangle^{}_{\mathrm{eff}} - 
    \langle \left( \hat{\mathcal{H}} 
    -\hat{\mathcal{H}}_{\mathrm{eff}} \right) \rangle^2_{\mathrm{eff}} \right] + ... \,. \nonumber
\end{eqnarray}
This expression is now an expansion in terms of cumulants, i.e. the $n$\textsuperscript{th} 
order of the expansion contains the effective expectation value of the $n$\textsuperscript{th} power 
of $\left(\hat{\mathcal{H}} -\hat{\mathcal{H}}_{\mathrm{eff}}\right)$ and the $n$\textsuperscript{th} power 
of the effective expectation value of $\left(\hat{\mathcal{H}} -\hat{\mathcal{H}}_{\mathrm{eff}}\right)$. 
The first non-trivial approximation of the free energy is obtained by cutting off the series after the 
first-order term. In order to obtain the original free energy, we have to set $\eta=1$, which leads to 
\begin{equation} \label{eq:F11approx}
  F^{(1)}(1) = F_{\mathrm{eff}} + \langle \left( \hat{\mathcal{H}} 
    -\hat{\mathcal{H}}_{\mathrm{eff}} \right) \rangle^{}_{\mathrm{eff}} \,.
\end{equation}
We can further evaluate the free energy $F^{(1)}(1)$ by inserting the original and the effective 
Hamiltonians~\eqeqref{eq:HamGen} and~\eqref{eq:Heff} and taking the effective expectation 
value~\eqref{eq:effexpval} of the occurring operators, to result in 
\begin{eqnarray}
  && F^{(1)}(1) = F_{\mathrm{eff}} + \sum_{\mb{n}} \sum_{\mb{n}'} \left[ E_{\mb{n},\mb{n}'} - 
    \left( \epsilon_{\mb{n}} - \mu \right) \delta_{\mb{n},\mb{n}'}
     \right] \,\langle \hat{a}^{\dag}_{\mb{n}} 
    \hat{a}^{}_{\mb{n}'} \rangle^{}_{\mathrm{eff}} \nonumber\\
   && + \frac{1}{2} 
    \sum_{\mb{n}} \sum_{\mb{m}} \sum_{\mb{m}'} \sum_{\mb{n}'} 
    U^{}_{\mb{n},\mb{m},\mb{m}',\mb{n}'} \,\langle \hat{a}^{\dag}_{\mb{n}} 
    \hat{a}^{\dag}_{\mb{m}} \hat{a}^{}_{\mb{m}'} \hat{a}^{}_{\mb{n}'} 
    \rangle^{}_{\mathrm{eff}} \,. 
\end{eqnarray}
We now process the effective expectation values further by applying the Wick rule \cite{1950Wick}. For 
the four-point correlation function in the interaction term, this leads to the decomposition 
into products of two-point correlation functions as 
\begin{eqnarray} \label{eq:Wick}
  \langle \hat{a}^{\dag}_{\mb{n}} \hat{a}^{\dag}_{\mb{m}} \hat{a}^{}_{\mb{m}'} \hat{a}^{}_{\mb{n}'} 
    \rangle^{}_{\mathrm{eff}} &=& \left( \delta_{\mb{n},\mb{n}'} \delta_{\mb{m},\mb{m}'} 
    + \delta_{\mb{n},\mb{m}'} \delta_{\mb{m},\mb{n}'} \right) \, \\
    && ~~~~~~~~~~~~ \times \langle \hat{a}^{\dag}_{\mb{n}} \hat{a}^{}_{\mb{n}} \rangle^{}_{\mathrm{eff}} \,
    \langle \hat{a}^{\dag}_{\mb{m}} \hat{a}^{}_{\mb{m}} \rangle^{}_{\mathrm{eff}} \,. \nonumber
\end{eqnarray}
From the investigation of the effective free energy, we can deduce a concrete expression 
for the two-point function that we are now left with. The effective free energy~\eqref{eq:Feff} reads 
with~\eqref{eq:Heff} and~\eqref{eq:Zeff} 
\begin{equation} \label{eq:FLHS}
  F_{\mathrm{eff}} = -\frac{1}{\beta} \ln \Tr \left[ e^{-\beta \sum_{\mb{n}} \left( 
    \epsilon_{\mb{n}} - \mu \right) \,\hat{a}^{\dag}_{\mb{n}} \hat{a}^{}_{\mb{n}} } \right]  \,,
\end{equation}
which reduces to 
\begin{equation} \label{eq:FRHS}
  F_{\mathrm{eff}} = \frac{1}{\beta} \sum_{\mb{n}} \ln \left[ 1- e^{-\beta \left( 
    \epsilon_{\mathrm{n}} - \mu \right) } \right]  \,.
\end{equation}
Differentiating both versions~\eqref{eq:FLHS},~\eqref{eq:FRHS} of $F_{\mathrm{eff}}$ with respect 
to the energies $\epsilon_{\mb{n}}$ leads to an identity for the expectation value of the 
two-point function, 
\begin{equation} \label{eq:ExpectValue}
  \langle \hat{a}^{\dag}_{\mb{n}} \hat{a}^{}_{\mb{n}} \rangle^{}_{\mathrm{eff}} = \frac{1}
    {e^{\beta \left( \epsilon_{\mathrm{n}} - \mu \right) } -1} \,,
\end{equation}
i.e. the Bose-Einstein distribution function. \\
We now introduce by hand the macroscopic occupation of the ground state, which is the predominant 
attribute of Bose-Einstein-condensation, by setting 
\begin{equation}
  \hat{a}^{\dag}_{\mb{0}} \simeq \hat{a}^{}_{\mb{0}} \simeq \sqrt{N_{\mb{0}}} = \psi \,,
\end{equation}
with $N_{\mb{0}}$ being the total number of particles in the ground state, which is characterized 
by the quantum number $\mb{n}=\mb{0}$. 
We now split all the terms into the $\mb{n}=\mb{0}$ and the $\mb{n}\neq \mb{0}$ 
contributions, and introduce a condensate wave function as 
\begin{equation}
  \Psi(\mb{x}) = \psi \, \Psi^{}_{\mb{0}}(\mb{x}),~~\Psi^{*}(\mb{x}) = \psi \, \Psi^{*}_{\mb{0}}(\mb{x}) \,.
\end{equation}
This wave function has the normalization 
\begin{equation} 
  \int d^3x\, \Psi^{*}_{}(\mb{x}) \Psi^{}_{}(\mb{x}) = 
    \psi^2 \int d^3x \,\Psi^{*}_{\mb{0}}(\mb{x}) \Psi^{}_{\mb{0}}(\mb{x}) = \psi^2 = N_{\mb{0}} \,.
\end{equation}
Note that the four-point correlation function as processed in~\eqeqref{eq:Wick} by the Wick rule, has to be 
modified for the condensate as 
\begin{equation}
  \langle \hat{a}^{\dag}_{\mb{0}} \hat{a}^{\dag}_{\mb{0}} \hat{a}^{}_{\mb{0}} \hat{a}^{}_{\mb{0}} 
    \rangle^{}_{\mathrm{eff}} = \psi^4 \,.
\end{equation}
Inserting the normalization~\eqref{eq:normcomm} for $\Psi^{}_{\mb{n}}(\mb{x})$ into the effective 
free energy, we have as a result 
\begin{eqnarray}
 F^{(1)}(1) = F_{\mathrm{eff}} + \left[ E_{\mb{0},\mb{0}} - \left( \epsilon_{\mb{0}} - \mu \right) \int d^3x \,
    \Psi^{*}_{\mb{0}}(\mb{x}) \Psi_{\mb{0}}(\mb{x}) \right] \, \psi^2 \nonumber\\
  + \sum_{\mb{n}\neq\mb{0}} \left[ E_{\mb{n},\mb{n}} - \left( \epsilon_{\mb{n}} - 
    \mu \right) \int d^3x \,\Psi^{*}_{\mb{n}}(\mb{x}) 
    \Psi_{\mb{n}}(\mb{x}) \right] \,\langle \hat{a}^{\dag}_{\mb{n}} 
    \hat{a}^{}_{\mb{n}} \rangle^{}_{\mathrm{eff}} ~~~~~ \nonumber\\ 
  + \frac{1}{2} \, U^{}_{\mb{0},\mb{0},\mb{0},\mb{0}} \, \psi^4 
    + \sum_{\mb{n}\neq \mb{0}} \left( U^{}_{\mb{n},\mb{0},\mb{0},\mb{n}} + 
    U^{}_{\mb{n},\mb{0},\mb{n},\mb{0}} \right) \,\psi^2 \,
    \langle \hat{a}^{\dag}_{\mb{n}} \hat{a}^{}_{\mb{n}} \rangle^{}_{\mathrm{eff}} ~~~~ \nonumber\\
  + \frac{1}{2} \,
    \sum_{\mb{n}\neq\mb{0}} \sum_{\mb{m}\neq\mb{0}} \left( U^{}_{\mb{n},\mb{m},\mb{m},\mb{n}} 
    + U^{}_{\mb{n},\mb{m},\mb{n},\mb{m}} \right) \,\langle \hat{a}^{\dag}_{\mb{n}} 
    \hat{a}^{}_{\mb{n}} \rangle^{}_{\mathrm{eff}} \,\langle \hat{a}^{\dag}_{\mb{m}} 
    \hat{a}^{}_{\mb{m}} \rangle^{}_{\mathrm{eff}} \,, \nonumber \\
  ~~~
\end{eqnarray}
where $F_{\mathrm{eff}}$ now consists of the two terms 
\begin{equation}
  F_{\mathrm{eff}} = (\epsilon_{\mb{0}} - \mu) \,\psi^2 +\frac{1}{\beta} \sum_{\mb{n}\neq\mb{0}} 
    \ln \left[ 1- e^{-\beta \left( \epsilon_{\mathrm{n}} - \mu \right) } \right]  \,.
\end{equation}
Inserting the expressions for the matrix elements $E_{\mb{n},\mb{n}'}$ and 
$U^{}_{\mb{n},\mb{m},\mb{m}',\mb{n}'}$ as defined in~\eqeqref{eq:Enn} and~\eqref{eq:Unmnm}, 
we can now write the total free energy, which will in the following be denoted shortly 
by $F$, as 
\begin{eqnarray} \label{eq:FtotalOrig}
  &&F = F_{\mathrm{eff}} + \int d^3x\, \Psi^{*}_{}(\mb{x}) \left[ h(\mb{x}) - 
    \mu \right] \Psi^{}_{}(\mb{x}) \\ 
  && ~~~~~~~~~~~ - \left( \epsilon_{\mb{0}} - \mu \right) \int d^3x\, 
    \Psi^{*}_{}(\mb{x}) \Psi^{}_{}(\mb{x}) \nonumber\\
  &&+ \sum_{\mb{n} \neq \mb{0}} \bigg\{ \int d^3x\, \Psi^{*}_{\mb{n}}(\mb{x}) \left[ h(\mb{x}) 
    - \mu \right] \Psi^{}_{\mb{n}}(\mb{x}) \nonumber\\
  && ~~~~~~~~~~~~~ - \left( \epsilon_{\mb{n}} - \mu \right) \int d^3x\, \Psi^{*}_{\mb{n}}(\mb{x}) 
    \Psi^{}_{\mb{n}}(\mb{x}) \bigg\} \langle \hat{a}^{+}_{\mb{n}} \hat{a}^{}_{\mb{n}} \rangle_{\mathrm{eff}} \nonumber\\
  &&+ \frac{1}{2} \int d^3x d^3x' \, \Psi^{*}_{}(\mb{x}) \Psi^{*}_{}(\mb{x}') \, U(\mb{x},\mb{x}') 
    \, \Psi^{}_{}(\mb{x}') \Psi^{}_{}(\mb{x}) \nonumber \\
  &&+ \sum_{\mb{n}\neq \mb{0}} \int d^3x d^3x' \, \Psi^{*}_{\mb{n}}(\mb{x}) \Psi^{*}_{}(\mb{x}') 
      \, U(\mb{x},\mb{x}') \nonumber\\
  && ~~~~~~~~~~~~ \times \Big[ \Psi^{}_{}(\mb{x}') \Psi^{}_{\mb{n}}(\mb{x}) + 
      \Psi^{}_{\mb{n}}(\mb{x}') \Psi^{}_{}(\mb{x}) \Big] \langle \hat{a}^{+}_{\mb{n}} 
      \hat{a}^{}_{\mb{n}} \rangle_{\mathrm{eff}} \nonumber \\
  &&+ \frac{1}{2} \,\sum_{\mb{n},\mb{m}\neq \mb{0}} \int d^3x d^3x' \, 
      \Big[ \Psi^{*}_{\mb{n}}(\mb{x}) \Psi^{*}_{\mb{m}}(\mb{x}') \,U(\mb{x},\mb{x}') \,
      \Psi^{}_{\mb{m}}(\mb{x}') \Psi^{}_{\mb{n}}(\mb{x}) \nonumber \\
  &&+ \Psi^{*}_{\mb{n}}(\mb{x}) \Psi^{*}_{\mb{m}}(\mb{x}') \,U(\mb{x},\mb{x}') 
      \,\Psi^{}_{\mb{n}}(\mb{x}') \Psi^{}_{\mb{m}}(\mb{x}) \Big] \langle \hat{a}^{+}_{\mb{n}} 
      \hat{a}^{}_{\mb{n}} \rangle_{\mathrm{eff}} \, \langle \hat{a}^{+}_{\mb{m}} \hat{a}^{}_{\mb{m}} 
      \rangle_{\mathrm{eff}} \,. \nonumber
\end{eqnarray}
In this theory, the condensate wave function encodes the behaviour of the particles in the 
condensate, i.e. a majority of particles in the system for low enough temperatures, while the 
wave functions with $\mb{n}\neq \mb{0}$ describe the thermal fluctuations on top of the 
condensate with increasing quantum numbers $\mb{n}$.

\subsection{Self-consistency equations}
As the unknown one-particle basis $\Psi^{}_{\mb{n}}(\mb{x})$ and energies $\epsilon_{\mb{n}}$ 
have been introduced artificially into the analysis, the result for the free energy should not 
depend on them. This is however only true for the exact expressions for $F$, and does not hold 
for the approximated form that we have used in the derivations following~\eqref{eq:F11approx}. 
This means that the approximation for the free energy $F$ does indeed depend on the one-particle basis and energies, but 
this dependence is unphysical and undesired. For this reason, we have to demand that the 
dependence of the free energy on these quantities be as small as possible - which mathematically 
corresponds to an extremization. This is the principle of minimal sensitivity, which was firstly 
introduced in Ref. \cite{1981Stev}. The equations obtained by varying the free energy~\eqref{eq:FtotalOrig} 
with respect to the condensate and thermal wave functions $\Psi^{*}(\mb{x})$ and $\Psi^{*}_{\mb{n}}(\mb{x})$, 
\begin{equation}
  \frac{\delta F}{\delta \Psi^{*}(\mb{x})} = \frac{\delta F}{\delta \Psi^{*}_{\mb{n}}(\mb{x})} = 0 \,.
\end{equation}
are called first and second Hartree-Fock equations, respectively. Furthermore, also the variation of the 
free energy with respect to the one-particle energies $\epsilon_{\mb{n}}$ must vanish, 
\begin{equation}
  \frac{\partial F}{\partial \epsilon_{\mb{n}}} = 0 \,.
\end{equation}
Finally, the derivation of $F$ with respect to the chemical potential must yield the total number 
of particles $N$ in the system, 
\begin{equation}
    -\frac{\partial F}{\partial\mu} = N \,.
\end{equation}
We now define the densities of condensate and thermal fluctuations as 
\begin{equation} \label{eq:densities}
	n_{\mb{0}}(\mb{x}) = |\Psi (\mb{x})|^2 \,,~~
	n_{\mathrm{th}}(\mb{x},\mb{x}') = \sum_{\mb{n}\neq \mb{0}}^{} \frac{\Psi^{*}_{\mb{n}}(\mb{x})\, \Psi^{~}_{\mb{n}}(\mb{x}')}
		{e^{\beta(\epsilon_{\mb{n}}-\mu)}-1} \,.
\end{equation}
For equal arguments of the thermal density, we will use the abbreviation 
$n_{\mathrm{th}}(\mb{x},\mb{x}) = n_{\mathrm{th}}(\mb{x})$. \\
With this, the variation of the free energy with respect to the condensate wave function leads to the 
first Hartree-Fock equation, 
\begin{eqnarray} \label{eq:HF1}
  \frac{\delta F}{\delta \Psi^{*}(\mb{x})} &=& \left[ h(\mb{x}) - \mu \right] \Psi(\mb{x}) \\
  && + \int d^3\mb{x}' \,U(\mb{x},\mb{x}') 
    \bigg\{ \big[ n_{\mb{0}}(\mb{x}') + n_{\mathrm{th}}(\mb{x}') \big] \Psi(\mb{x}) \nonumber\\
  && ~~~~~~~~~~~~~~~~~~~~~ + n_{\mathrm{th}}(\mb{x}',\mb{x}) \Psi(\mb{x}') \bigg\} =0 \,, \nonumber
\end{eqnarray}
whereas the variation of $F$ with respect to the thermal wave functions yields the second Hartree-Fock equation, 
\begin{eqnarray} \label{eq:HF2}
    \frac{\delta F}{\delta \Psi^{*}_{\mb{n}}(\mb{x})} &=& \left[ h(\mb{x}) - \epsilon_{\mb{n}} \right] 
    \Psi^{}_{\mb{n}}(\mb{x}) \\
    && + \int d^3\mb{x}' \,U(\mb{x},\mb{x}') \bigg\{ \big[ n_{\mb{0}}(\mb{x}') 
    + n_{\mathrm{th}}(\mb{x}') \big] \Psi^{}_{\mb{n}}(\mb{x}) \nonumber \\
    &&~~~~~~
    + \big[ \Psi^{*}(\mb{x}') \Psi(\mb{x}) + n_{\mathrm{th}}(\mb{x}',\mb{x}) \big] 
    \Psi^{}_{\mb{n}}(\mb{x}') \bigg\} =0 \,. \nonumber
\end{eqnarray}
In both equations, the first, local part of the interaction is referred to as Hartree term, 
or  direct interaction term, whereas the second, bilocal part is the Fock term, or exchange 
interaction term. \\
The derivation of the free energy with respect to the energies $\epsilon_{\mb{n}}$ reproduces 
the already known identity~\eqref{eq:ExpectValue} for the expectation value of the two-point 
correlation function of the creation and annihilation operators. Finally, the negative derivative 
of the free energy with respect to the chemical potential, 
\begin{equation} \label{eq:partN}
  N= \int d^3x\,\left[ n_{\mb{0}}(\mb{x}) + n_{\mathrm{th}}(\mb{x}) \right] \,
\end{equation}
recovers correctly the total number of particles in the system.

\subsection{Semi-classical limit}
Instead of using the wave functions of condensate and thermal fluctuations, we now 
pursue a different approach and define the densities of condensate and thermal cloud 
in the semi-classical limit as the basic variables instead. Let us thus first take 
the semi-classical limit of the free energy, introducing both condensate and thermal 
density instead of the wave functions, and then show that it is possible to derive the correct 
Hartree-Fock equations by variation of the semi-classical free energy with respect to the 
respective densities. \\
In the semi-classical approximation we use plane waves as an ansatz for the 
thermal wave functions, i.e. $\Psi^{}_{\mb{n}}(\mb{x}) \rightarrow 
\Psi^{}_{\mb{k}}(\mb{x}) = e^{i \mb{k} \mb{x}}$, so the discrete energies $\epsilon_{\mb{n}}$ 
become local dispersions $\epsilon_{\mb{k}}(\mb{x})$. Furthermore, we apply the Thomas-Fermi 
approximation for the condensate, which means neglecting the Laplace term for the condensate 
wave functions. In addition, the sums over the quantum numbers $\mb{n}$ are replaced by  
integrals in $\mb{k}$-space, which changes the thermal density in~\eqref{eq:densities} to 
\begin{equation} \label{eq:nthSC}
    n_{\mathrm{th}}(\mb{x}) = \int \frac{d^3k}{(2\pi)^3} \, n_{\mathrm{th}}(\mb{x},\mb{k}) \,.
\end{equation}
Here we have defined the thermal Wigner quasiprobability, 
\begin{equation}
  n_{\mathrm{th}}(\mb{x},\mb{k}) = \frac{1}{e^{\beta \left[ \epsilon_{\mb{k}}(\mb{x}) -\mu \right]}} \,,
\end{equation}
which will become the variational parameter instead of the thermal density itself. Applying all the 
prescriptions above, the semi-classical approximation of the free energy~\eqref{eq:FtotalOrig} reads 
\begin{eqnarray} \label{eq:Fsc}
  && F_{\mathrm{SC}} = \frac{1}{\beta} \int d^3x\, \int \frac{d^3k}{(2\pi)^3} \,
    \ln \left\{ 1-e^{-\beta \left[\epsilon_{\mb{k}}(\mb{x})-\mu\right]} \right\} \\
  && + \int d^3x \, \left[ V(\mb{x}) - \mu \right] \, n_{\mb{0}}(\mb{x}) \nonumber \\
  && + \int d^3x \, \int \frac{d^3k}{(2\pi)^3} \, \left[ \frac{\hbar^2 \mb{k}^2}{2m} 
    + V(\mb{x}) - \epsilon_{\mb{k}}(\mb{x}) \right] n_{\mathrm{th}}(\mb{x},\mb{k}) \nonumber \\
  && + \int d^3x d^3x' \, U(\mb{x},\mb{x}') \,\Big[ \frac{1}{2} \,n_{\mb{0}}(\mb{x})\,
    n_{\mb{0}}(\mb{x}') + n_{\mb{0}}(\mb{x}')\,n_{\mathrm{th}}(\mb{x}) \nonumber \\
  && + \sqrt{n_{\mb{0}}(\mb{x}') n_{\mb{0}}(\mb{x})} \, n_{\mathrm{th}}(\mb{x},\mb{x}') 
    + \frac{1}{2} \,n_{\mathrm{th}}(\mb{x}) \, n_{\mathrm{th}}(\mb{x}') \nonumber\\
  && ~~~~~~~~~ + \frac{1}{2} n_{\mathrm{th}}(\mb{x},\mb{x}') \, n_{\mathrm{th}}(\mb{x}',\mb{x}) \Big] \,.\nonumber
\end{eqnarray}
Let us now derive the semi-classical Hartree-Fock equations by variation of $F_{\mathrm{SC}}$ with respect to the 
densities. The extremization of $F_{\mathrm{SC}}$ with respect to the condensate density $n_{\mb{0}}(\mb{x})$ yields 
  \begin{eqnarray} \label{eq:HF1sc}
    \frac{\delta F_{\mathrm{SC}}}{\delta n_{\mb{0}}(\mb{x})} = V(\mb{x}) - \mu + 
      \int d^3x' \, U(\mb{x},\mb{x}') \,\Big[ n_{\mb{0}}(\mb{x}') + 
      n_{\mathrm{th}}(\mb{x}') \Big] \nonumber\\ 
    + \frac{1}{2} \,\int d^3x' \, U(\mb{x},\mb{x}') \,
      \sqrt{\frac{n_{\mb{0}}(\mb{x}')}{n_{\mb{0}}(\mb{x})}} \, 
      \Big[ n_{\mathrm{th}}(\mb{x}',\mb{x}) + n_{\mathrm{th}}(\mb{x},\mb{x}') \Big] =0 \,.\nonumber \\
    ~~~ 
  \end{eqnarray}
  Considering a multiplication with the wave function $\Psi(\mb{x}) \equiv \sqrt{n_{\mb{0}}(\mb{x})}$, 
  this correctly corresponds to the Hartree-Fock equation for the condensate~\eqref{eq:HF1} in the 
  Thomas-Fermi-approximation. \\
  In view of the variation of $F_{\mathrm{SC}}$ with respect to the thermal quasiprobability $n_{\mathrm{th}}(\mb{x},\mb{k})$
  we first have to define the Wigner quasiprobability function 
  for the bilocal thermal density. Generalizing the notion~\eqref{eq:nthSC} 
  straightforwardly for different arguments $\mb{x},\mb{x}'$, we get from~\eqref{eq:densities} 
  \begin{eqnarray} 
    n_{\mathrm{th}}(\mb{x},\mb{x}') \rightarrow n_{\mathrm{th}}(\mb{R},\mb{s}) &=& \int \frac{d^3k}{(2\pi)^3} \, 
      \frac{e^{-i\mb{k} \mb{s}}}{e^{\beta \left[ \epsilon_{\mb{k}}(\mb{R}) - 
      \mu \right]} -1} \nonumber\\
    &=:& \int \frac{d^3k}{(2\pi)^3} \, e^{-i\mb{k}\mb{s}} \,
      n_{\mathrm{th}}(\mb{R},\mb{s},\mb{k}) \,,\nonumber \\
    ~~~ 
  \end{eqnarray}
  where we have adapted the center-of-mass coordinate $\mb{R} = (\mb{x}+\mb{x}')/2$ and 
  the relative coordinate $\mb{s}=\mb{x}-\mb{x}'$ instead of $\mb{x}$ and $\mb{x}'$. 
  This general definition is in accordance with the definition~\eqref{eq:nthSC} for the local 
  expression of the thermal density, since in the case $\mb{x}=\mb{x}'$ we have 
  \begin{eqnarray} 
    n_{\mathrm{th}}(\mb{x},\mb{x}) \equiv n_{\mathrm{th}}(\mb{R},\mb{s}=\mb{0}) &=& 
      \int \frac{d^3k}{(2\pi)^3} \, \frac{1}{e^{\beta \left[ \epsilon_{\mb{k}}(\mb{R}) - 
      \mu \right]} -1} \nonumber\\
    &=& \int \frac{d^3k}{(2\pi)^3} \, n_{\mathrm{th}}(\mb{R},\mb{k}) \,,
  \end{eqnarray}
  which is identical with the Wigner quasiprobability defined in~\eqeqref{eq:nthSC}. 
  The semi-classical free energy~\eqref{eq:Fsc} can be rewritten in terms of $\mb{R}$ and $\mb{s}$ as 
  \begin{eqnarray} \label{eq:FscRs}
    && F_{\mathrm{SC}} = \frac{1}{\beta} \int d^3R\, \int \frac{d^3k}{(2\pi)^3} \,
      \ln \left\{ 1-e^{-\beta \left[\epsilon_{\mb{k}}(\mb{R})-\mu\right]} \right\} \\
    && + \int d^3R \, \left[ V(\mb{R}) - \mu \right] \, n_{\mb{0}}(\mb{R}) \nonumber \\
    && + \,\frac{1}{2} \int d^3R d^3s \, U(\mb{s}) \, n_{\mb{0}}\left(\mb{R}+\frac{\mb{s}}{2}\right)\,
      n_{\mb{0}}\left(\mb{R}-\frac{\mb{s}}{2}\right) \nonumber\\
    && + \int d^3R \, \int \frac{d^3k}{(2\pi)^3} \, \left[ \frac{\hbar^2 \mb{k}^2}{2m} 
      + V(\mb{R}) - \epsilon_{\mb{k}}(\mb{R}) \right] n_{\mathrm{th}}(\mb{R},\mb{k}) \nonumber \\
    && + \int d^3R d^3s \int \frac{d^3k}{(2\pi)^3} \, U(\mb{s}) \, \bigg[ 
      n_{\mb{0}}\left(\mb{R}-\frac{\mb{s}}{2}\right)
      \,n_{\mathrm{th}}\left(\mb{R}+\frac{\mb{s}}{2}\right) \nonumber\\
    && ~~~~~~~ + \sqrt{n_{\mb{0}}\left(\mb{R}+\frac{\mb{s}}{2}\right) 
      n_{\mb{0}}\left(\mb{R}-\frac{\mb{s}}{2}\right)} 
      ~ n_{\mathrm{th}}(\mb{R},\mb{k}) ~ e^{-i \mb{k} \mb{s}} \bigg] \nonumber\\
    && + \,\frac{1}{2} \int d^3R d^3s \int \frac{d^3k d^3k'}{(2\pi)^6} \, U(\mb{s}) \nonumber\\
    && ~~~~~~~ \times \bigg[ n_{\mathrm{th}}\left(\mb{R}+\frac{\mb{s}}{2},\mb{k}\right) \, 
      n_{\mathrm{th}}\left(\mb{R}-\frac{\mb{s}}{2},\mb{k}'\right) \nonumber\\
    && ~~~~~~~~~~~~~~ + e^{-i \mb{s} (\mb{k} - \mb{k}')} \, 
      n_{\mathrm{th}}(\mb{R},\mb{k}) \, n_{\mathrm{th}}(\mb{R},\mb{k}') \bigg] \,. \nonumber 
  \end{eqnarray}
  The total variation of the free energy with respect to $n_{\mathrm{th}}(\mb{R},\mb{k})$ then 
  reads 
  \begin{eqnarray}
   && \frac{\delta F_{\mathrm{SC}}}{\delta n_{\mathrm{th}}(\mb{R},\mb{k})} = 
      \frac{\hbar^2 \mb{k}^2}{2m} + V(\mb{R}) + \int d^3s \, 
      U(\mb{s}) \, n_{\mb{0}}\left(\mb{R}-\frac{\mb{s}}{2}\right) \nonumber\\
   && - \epsilon_{\mb{k}}(\mb{R}) + \int d^3s \, U(\mb{s}) \,e^{-i \mb{k} \mb{s}} \sqrt{n_{\mb{0}}\left(\mb{R} + 
      \frac{\mb{s}}{2}\right) n_{\mb{0}}\left(\mb{R}-\frac{\mb{s}}{2}\right)} \nonumber \\
   && + \int d^3s \int \frac{d^3k'}{(2\pi)^3} \, U(\mb{s})
      \bigg[ n_{\mathrm{th}}\left(\mb{R}-\frac{\mb{s}}{2},\mb{k}'\right) \nonumber\\
   && ~~~~~~~~~~~~~~~~~~ + e^{-i \mb{s} (\mb{k}-\mb{k}')} \,
      n_{\mathrm{th}}\left(\mb{R}+\frac{\mb{s}}{2},\mb{k}\right) \bigg] =0\,.
  \end{eqnarray}
  This yields the local dispersion of the thermal fluctuations, now in terms of $\mb{x}$ and $\mb{x}'$, as 
  \begin{eqnarray} \label{eq:ekxx'}
    \epsilon_{\mb{k}}(\mb{x}) &=& \frac{\hbar^2 \mb{k}^2}{2m} + V(\mb{x}) + \int d^3x' \, 
      U(\mb{x},\mb{x}') \, n_{\mb{0}}(\mb{x}')  \\
    &&  + \int d^3x' \,  U(\mb{x},\mb{x}') \,e^{-i \mb{k} (\mb{x}-\mb{x}')} 
      \sqrt{n_{\mb{0}}\left(\mb{x}\right) n_{\mb{0}}\left(\mb{x}'\right)} \nonumber\\
    && + \int d^3x' \int \frac{d^3k'}{(2\pi)^3} \,  U(\mb{x},\mb{x}')
      \bigg[ n_{\mathrm{th}}(\mb{x}',\mb{k}') \nonumber\\
    && ~~~~~~~~~~~ + e^{-i (\mb{x}-\mb{x}') (\mb{k}-\mb{k}')} \,
      n_{\mathrm{th}}(\mb{x},\mb{k}') \bigg] \,. \nonumber
  \end{eqnarray}
  The derivation of $F_{\mathrm{SC}}$ with respect to the energies $\epsilon_{\mb{k}}(\mb{x})$ 
  simply rederives the form of the function $n_{\mathrm{th}}(\mb{x},\mb{k})$ as 
  introduced in~\eqeqref{eq:nthSC}. 
  The derivation of $F_{\mathrm{SC}}$ with respect to the chemical potential $\mu$ leads as expected again 
to the particle number equation~\eqref{eq:partN}. 
The fact that we obtained consistent equations from the variation of the semi-classical free 
energy with respect to the condensate and thermal density shows that the semi-classical limit 
conserves the physical properties of the system. 
The semi-classical output of the Hartree-Fock theory consists thus of the equation of motion for 
the condensate density \eqeqref{eq:HF1sc} and the semi-classical energies of the thermal 
fluctuations \eqeqref{eq:ekxx'}. 

\subsection{Specializing to contact and gravitational interaction}
These two Hartree-Fock equations can now be specified to a system with repulsive contact and attractive 
gravitational interactions. To this end the general interaction $U(\mb{x}-\mb{x}')$ is replaced by~\eqeqref{eq:interactions}, 
and we set the external potential to zero, i.e. $V(\mb{x})=0$. In terms of these interactions, the 
two exact Hartree-Fock equations~\eqref{eq:HF1} and~\eqref{eq:HF2} read 
\begin{eqnarray} \label{eq:HF1exact}
  && - \mu \, \Psi(\mb{x}) + g\, \left[ n_{\mb{0}}(\mb{x}) + 
    2n_{\mathrm{th}}(\mb{x}) \right] \Psi(\mb{x}) \\[8pt]
  && ~~~~~~~  - \int d^3\mb{x}' \, \frac{G m^2}{|\mb{x}-\mb{x}'|}
    \bigg\{ \left[ n_{\mb{0}}(\mb{x}') + n_{\mathrm{th}}(\mb{x}') \right] \Psi(\mb{x}) \nonumber\\ 
  && ~~~~~~~~~~~~~~~~ + n_{\mathrm{th}}(\mb{x}',\mb{x}) \Psi(\mb{x}') \bigg\} = 0 \,,\nonumber
\end{eqnarray}
and 
\begin{eqnarray} \label{eq:HF2exact}
  &&\left[ h(\mb{x}) - \epsilon_{\mb{n}} \right] \Psi^{}_{\mb{n}}(\mb{x}) + 
    g \, \left[ 2 n_{\mb{0}}(\mb{x}) + 
    2n_{\mathrm{th}} (\mb{x}) \right] \Psi^{}_{\mb{n}}(\mb{x}) \\[8pt]
  && ~~~~~ - \int d^3\mb{x}' \, \frac{G m^2}{|\mb{x}-\mb{x}'|} \bigg\{ \left[ n_{\mb{0}}(\mb{x}') + 
    n_{\mathrm{th}}(\mb{x}') \right] \Psi^{}_{\mb{n}}(\mb{x}) \nonumber \\
  && ~~~~~~~ + \left[ \sqrt{n_{\mb{0}}(\mb{x}') n_{\mb{0}}(\mb{x})} + 
    n_{\mathrm{th}}(\mb{x}',\mb{x}) \right] \Psi^{}_{\mb{n}}(\mb{x}') \bigg\} = 0 \,, \nonumber
\end{eqnarray}
respectively, where we have used the identification 
\begin{equation}
	\Psi^{*}(\mb{x}') \Psi(\mb{x}) \equiv \sqrt{n_{\mb{0}}(\mb{x}') n_{\mb{0}}(\mb{x})} \,
\end{equation}
by assuming that the condensate wave function $\Psi(\mb{x})$ just contains a global phase, which is 
justified for a stationary superfluid with vanishing velocity. 
The first parts of each equation are familiar from a system of particles in an external trap considering 
only contact interaction between the particles, as is the case for most BEC experiments in the lab. 
With the gravitational interaction the situation becomes less convenient due to its nonlocality. 
In particular, the Fock terms of the gravitational interaction pose a problem since they contain the bilocal 
form of the respective densities, i.e. $\sqrt{n_{\mb{0}}(\mb{x}') n_{\mb{0}}(\mb{x})}$ and 
$n_{\mathrm{th}}(\mb{x}',\mb{x})$. Due to the mathematical difficulties related to these terms, we discard the 
bilocal contributions to the theory, i.e. we will carry out a Hartree-approximation 
for the gravitational interaction, ad neglect the bilocal Fock-terms. \\
With this we conclude from~\eqeqref{eq:HF1exact} that the equation is fulfilled either if the wave function 
$\Psi(\mb{x})$ is zero, or the equation 
\begin{eqnarray} \label{eq:HF1CG}
  && - \mu + g\,\left[ n_{\mb{0}}(\mb{x}) + 2\, n_{\mathrm{th}}(\mb{x}) \right] \\
  && ~~~~~~~~~~~~~ - \int d^3x' \, \frac{G m^2}{|\mb{x}-\mb{x}'|} 
      \, \Bigg[ n_{\mb{0}}(\mb{x}') + n_{\mathrm{th}}(\mb{x}') \Bigg] =0 \, \nonumber
\end{eqnarray}
holds. Therefore, the system contains two regimes - one, where the condensate wave function, or density, 
vanishes, and another, in which the condensate density is non-zero and the dynamics of condensate and thermal 
density are determined by~\eqeqref{eq:HF1CG}. \\
The semi-classical limit of the second Hartree-Fock equation~\eqref{eq:HF2exact} just gives the thermal 
energies~\eqref{eq:ekxx'} as a function of the wavenumber $\bf{k}$, 
\begin{eqnarray} \label{eq:ekCG}
   && \epsilon_{\mb{k}}(\mb{x}) = \frac{\hbar^2 \mb{k}^2}{2m} 
      + 2 g\, \big[ n_{\mb{0}}(\mb{x}) + n_{\mathrm{th}}(\mb{x}) \big] \\
  && ~~~~~~~ - \int d^3x' \, \frac{G m^2}{|\mb{x}-\mb{x}'|} \, 
    \bigg[n_{\mb{0}}(\mb{x}') + n_{\mathrm{th}}(\mb{x}') \bigg] \,. \nonumber
\end{eqnarray}
The two equations~\eqref{eq:HF1CG} and~\eqref{eq:ekCG} will be the starting point of our calculations in~\secref{sec:CG}.

\end{document}